\documentclass[apj,numberedappendix,revtex4,preprint]{emulateapj}

\usepackage{amsmath,amsbsy,apjfonts,bm,color,graphics,multirow,footmisc,hyperref,Macro}
\hypersetup{
  colorlinks = true,
  urlcolor   = blue,
  linkcolor  = blue,
  citecolor  = blue
}

\def\bicep{{\sc BICEP}}
\def\bicepone{{\sc BICEP1}}
\def\biceptwo{{\sc BICEP2}}
\def\keckarray{{\it Keck Array}}
\def\keck{{\it Keck}}
\def\BK{\bicep\ / \keck}
\def\bk{{\rm BK}14}
\def\planck{{\it Planck}}

\def\sptpol{{\sc SPTpol}}
\def\act{{\sc ACT}}
\def\spt{{\sc SPT}}
\def\sptIIIg{{\sc SPT3G}}
\def\polarbear{{\sc Polarbear}}

\def\grad{\phi}
\def\curl{\varpi}
\def\estg{\widehat{\grad}}
\def\estc{\widehat{\curl}}
\def\uestg{\ol{\grad}}

\def\hckk{\widehat{C}^{\grad\grad}}
\def\AL{A_{\rm L}^{\grad\grad}}
\def\ALB{A_{\rm L}^{\rm BB}}

\def\bE{\ol{E}}
\def\bB{\ol{B}}
\def\bX{\ol{X}}
\def\bY{\ol{Y}}
\def\tE{\widetilde{E}}
\def\tB{\widetilde{B}}
\def\hE{\widehat{E}}
\def\hB{\widehat{B}}
\def\hX{\widehat{X}}
\def\hC{\widehat{C}}
\def\hN{\widehat{N}}
\def\tCEE{\widetilde{C}^{\rm EE}}

\def\hCEE{\hC^{\rm EE}}
\def\hCBB{\hC^{\rm BB}}
\def\hCXX{\hC^{\rm XX}}

\def\ol{\overline}
\def\l{\ell}
\def\lmax{\ell_{\rm max}}
\def\lmin{\ell_{\rm min}}
\def\bl{\bm{\ell}}
\def\bn{\bm{\nabla}}
\def\bL{\textit{\textbf{L}}}
\def\bd{\textit{\textbf{d}}}
\def\hatn{\hat{\textit{\textbf{n}}}}
\def\fig#1{Fig.~\ref{#1}}

\def\sec#1{Sec.~\ref{#1}}
\def\eq#1{Eq.~\eqref{#1}}

\def\mALc{1.13}
\def\vALc{0.20}
\def\mALa{1.15}
\def\vALa{0.36}
\def\sULa{5.8}
\def\mALb{1.20}
\def\vALb{0.17}

\def\email#1{\href{mailto:#1}{#1}}

\shorttitle{Measurement of gravitational lensing from large-scale $B$-mode polarization}
\shortauthors{\keckarray\ and \biceptwo\ Collaborations}
\submitted{to be submitted to \apj --- draft version \today}

\begin{document}

% Title %
\title{\biceptwo\ / \keckarray\ VIII: Measurement of gravitational lensing from large-scale $B$-mode polarization}

% Authors %
\author{\keckarray\ and \biceptwo\ Collaborations: 
P.~A.~R.~Ade\altaffilmark{1}}
\author{Z.~Ahmed\altaffilmark{2,3}}
\author{R.~W.~Aikin\altaffilmark{4}}
\author{K.~D.~Alexander\altaffilmark{5}}
\author{D.~Barkats\altaffilmark{5}}
\author{S.~J.~Benton\altaffilmark{6}}
\author{C.~A.~Bischoff\altaffilmark{5}}
\author{J.~J.~Bock\altaffilmark{7,8}}
\author{R.~Bowens-Rubin\altaffilmark{5}}
\author{J.~A.~Brevik\altaffilmark{7}}
\author{I.~Buder\altaffilmark{5}}
\author{E.~Bullock\altaffilmark{9}}
\author{V.~Buza\altaffilmark{5,10}}
\author{J.~Connors\altaffilmark{5}}
\author{B.~P.~Crill\altaffilmark{8}}
\author{L.~Duband\altaffilmark{11}}
\author{C.~Dvorkin\altaffilmark{10}}
\author{J.~P.~Filippini\altaffilmark{7,12}}
\author{S.~Fliescher\altaffilmark{9}}
\author{J.~Grayson\altaffilmark{2}}
\author{M.~Halpern\altaffilmark{13}}
\author{S.~Harrison\altaffilmark{5}}
\author{S.~R.~Hildebrandt\altaffilmark{7,8}}
\author{G.~C.~Hilton\altaffilmark{14}}
\author{H.~Hui\altaffilmark{7}}
\author{K.~D.~Irwin\altaffilmark{2,3}}
\author{J.~Kang\altaffilmark{2,3}}
\author{K.~S.~Karkare\altaffilmark{5}}
\author{E.~Karpel\altaffilmark{2}}
\author{J.~P.~Kaufman\altaffilmark{15}}
\author{B.~G.~Keating\altaffilmark{15}}
\author{S.~Kefeli\altaffilmark{7}}
\author{S.~A.~Kernasovskiy\altaffilmark{1}}
\author{J.~M.~Kovac\altaffilmark{5,10}}
\author{C.~L.~Kuo\altaffilmark{2,3}}
\author{E.~M.~Leitch\altaffilmark{16}}
\author{M.~Lueker\altaffilmark{7}}
\author{K.~G.~Megerian\altaffilmark{8}}
\author{T.~Namikawa\altaffilmark{2,3,$\dagger$}}
\author{C.~B.~Netterfield\altaffilmark{6,17}}
\author{H.~T.~Nguyen\altaffilmark{8}}
\author{R.~O'Brient\altaffilmark{7,8}}
\author{R.~W.~Ogburn~IV\altaffilmark{2,3}}
\author{A.~Orlando\altaffilmark{7}}
\author{C.~Pryke\altaffilmark{9,18}}
\author{S.~Richter\altaffilmark{5}}
\author{R.~Schwarz\altaffilmark{18}}
\author{C.~D.~Sheehy\altaffilmark{16,18}}
\author{Z.~K.~Staniszewski\altaffilmark{7,8}}
\author{B.~Steinbach\altaffilmark{7}}
\author{R.~V.~Sudiwala\altaffilmark{1}}
\author{G.~P.~Teply\altaffilmark{7,15}}
\author{K.~L.~Thompson\altaffilmark{2,3}}
\author{J.~E.~Tolan\altaffilmark{2}}
\author{C.~Tucker\altaffilmark{1}}
\author{A.~D.~Turner\altaffilmark{8}}
\author{A.~G.~Vieregg\altaffilmark{5,16,19}}
\author{A.~C.~Weber\altaffilmark{8}}
\author{D.~V.~Wiebe\altaffilmark{13}}
\author{J.~Willmert\altaffilmark{18}}
\author{C.~L.~Wong\altaffilmark{5,10}}
\author{W.~L.~K.~Wu\altaffilmark{2,20}}
\author{K.~W.~Yoon\altaffilmark{2}}

\altaffiltext{1}{School of Physics and Astronomy, Cardiff University, Cardiff, CF24 3AA, United Kingdom}
\altaffiltext{2}{Department of Physics, Stanford University, Stanford, CA 94305, USA}
\altaffiltext{3}{Kavli Institute for Particle Astrophysics and Cosmology, SLAC National Accelerator Laboratory, 2575 Sand Hill Rd, Menlo Park, CA 94025, USA}
\altaffiltext{4}{Department of Physics, California Institute of Technology, Pasadena, California 91125, USA}
\altaffiltext{5}{Harvard-Smithsonian Center for Astrophysics, 60 Garden Street MS 42, Cambridge, Massachusetts 02138, USA}
\altaffiltext{6}{Department of Physics, University of Toronto, Toronto, Ontario, M5S 1A7, Canada}
\altaffiltext{7}{Department of Physics, California Institute of Technology, Pasadena, California 91125, USA}
\altaffiltext{8}{Jet Propulsion Laboratory, Pasadena, California 91109, USA}
\altaffiltext{9}{Minnesota Institute for Astrophysics, University of Minnesota, Minneapolis, Minnesota 55455, USA}
\altaffiltext{10}{Department of Physics, Harvard University, Cambridge, MA 02138, USA}
\altaffiltext{11}{Service des Basses Temp\'{e}ratures, Commissariat \`{a} l'Energie Atomique, 38054 Grenoble, France}
\altaffiltext{12}{Department of Physics, University of Illinois at Urbana-Champaign, Urbana, Illinois 61801, USA}
\altaffiltext{13}{Department of Physics and Astronomy, University of British Columbia, Vancouver, British Columbia, V6T 1Z1, Canada}
\altaffiltext{14}{National Institute of Standards and Technology, Boulder, Colorado 80305, USA}
\altaffiltext{15}{Department of Physics, University of California at San Diego, La Jolla, California 92093, USA}
\altaffiltext{16}{Kavli Institute for Cosmological Physics, University of Chicago, Chicago, IL 60637, USA} 
\altaffiltext{17}{Canadian Institute for Advanced Research, Toronto, Ontario, M5G 1Z8, Canada}
\altaffiltext{18}{School of Physics and Astronomy, University of Minnesota, Minneapolis, Minnesota 55455, USA}
\altaffiltext{19}{Department of Physics, Enrico Fermi Institute, University of Chicago, Chicago, IL 60637, USA}
\altaffiltext{20}{Department of Physics, University of California, Berkeley, CA 94720, USA}
\altaffiltext{$\dagger$}{Corresponding author: T.~Namikawa, \email{namikawa@slac.stanford.edu}}

% Abstract %
%----------------------------------------------------------------------------------------------------%
\begin{abstract}

We present measurements of polarization lensing using the 150\,GHz maps which include
all data taken by the \biceptwo\ \& \keckarray\ CMB polarization experiments
up to and including the 2014 observing season (\bk).
Despite their modest angular resolution ($\sim 0.5^\circ$), the excellent sensitivity 
($\sim 3\mu$K-arcmin) of these maps makes it possible to directly reconstruct the lensing potential using 
only information at larger angular scales ($\l\leq 700$).
From the auto-spectrum of the reconstructed potential we measure an amplitude of the spectrum 
to be $\AL=\mALa\pm\vALa$ (\planck\ $\Lambda$CDM prediction corresponds to $\AL=1$), 
and reject the no-lensing hypothesis at $\sULa\sigma$, 
which is the highest significance achieved to date using an $EB$ lensing estimator.
Taking the cross-spectrum of the reconstructed potential with
the \planck\ 2015 lensing map yields $\AL=\mALc\pm\vALc$.
These direct measurements of $\AL$ are consistent with the $\Lambda$CDM cosmology, and
with that derived from the previously reported \bk\ $B$-mode auto-spectrum ($\ALB=\mALb\pm\vALb$).
We perform a series of null tests and consistency checks to show that these results are robust against 
systematics and are insensitive to analysis choices.
These results unambiguously demonstrate that the $B$-modes previously reported by \BK\ 
at intermediate angular scales ($150\lesssim\l\lesssim350$) are dominated by gravitational lensing. 
The good agreement between the lensing amplitudes obtained from the lensing reconstruction and $B$-mode 
spectrum starts to place constraints on any alternative cosmological sources of $B$-modes 
at these angular scales. 
\end{abstract}
%----------------------------------------------------------------------------------------------------%

%////////////////////////////////////////////////////////////////////////////////////////////////////%
% MAIN MATTER 
%////////////////////////////////////////////////////////////////////////////////////////////////////%

\maketitle

% Contents %
%////////////////////////////////////////////////////////////////////////////////////////////////////%
\section{Introduction} \label{sec.1}
%////////////////////////////////////////////////////////////////////////////////////////////////////%

Cosmic Microwave Background (CMB) photons traveling from the surface of last scattering are
lensed by the gravitational potential of the large-scale structure along the line of sight. 
This leads to spatial distortions of a few arcminutes in the temperature and polarization anisotropies. 
In particular, gravitational lensing converts some of the $E$-mode polarization into
$B$-mode polarization \citep{Zaldarriaga:1998ar}.
Measurements of temperature and polarization 
with sufficient resolution and sensitivity can be used to reconstruct the intervening matter distribution, 
and in the future such bias-free measurements of large-scale structure will become some of the
most powerful probes in cosmology
(e.g., \citealt{Hu:2001a,Namikawa:2011,Wu:2014,Abazajian:2013oma,Allison:2015,Pan:2015}).
Lensing can also act as a noise source for primordial $B$-modes, which peak at degree-scales 
(e.g., \citealt{Kesden:2002,Knox:2002}). 
With sufficient sensitivity, a reconstructed lensing potential can be used to predict the degree-scale 
lensing $B$-modes, enabling a deeper search for a primordial signal.
If the tensor-to-scalar ratio $r$ is below $0.01$, such ``de-lensing'' procedures 
will become important in the search for inflationary $B$-modes 
\citep{Kesden:2002,Knox:2002,Seljak:2003pn,Smith:2010gu}.
In the latest \BK\ results we already see a non-negligible lensing contribution 
at large angular scales ($\l<100$) \citep{BKVI}.

Lensing reconstruction from high resolution CMB temperature maps has been performed using data from
the Atacama Cosmology Telescope (\act; \citealt{Das:2011,Das:2014}), \planck\ \citep{P13:phi} and 
the South Pole Telescope (\spt; \citealt{vanEngelen:2012,Story:2014hni}). 
More recently, reconstruction using polarization maps has also been demonstrated.
Using polarization data, the estimated amplitude of the lensing potential power spectrum, $\AL$, from 
\planck\ 2015, \polarbear\ and \sptpol\ are, $\AL=0.76\pm 0.15$ \citep{P15:phi}, 
$\AL=1.06\pm 0.47$ \citep{PB14:phi}, and 
$\AL=0.92\pm 0.24$ \citep{Story:2014hni}, respectively, 
where the errors denote the $1\,\sigma$ statistical uncertainties. 
The reconstructed lensing potential from the polarization maps can be used in cross-correlation with other 
lensing potential tracers such as 
the cosmic-infrared background (CIB) \citep{Hanson:2013daa,PB14:phixCIB,vanEngelen:2014zlh}. 
These measurements all use the fact that a common lensing potential introduces statistical
anisotropy into the observed CMB in the form of a correlation between the CMB polarization
anisotropies and their spatial derivatives \citep{Hu:2001,Hu:2001kj,Hirata:2003,Hirata:2003ka}.
These experiments have high enough angular resolution to resolve small-scale (arcminute) polarization fluctuations 
where weak lensing significantly perturbs the primordial CMB anisotropies.

The \biceptwo\ and \keckarray\ telescopes, with smaller apertures and beam sizes of $\sim 0.5^\circ$ at $150$\,GHz,
do not resolve the arcminute-scale fluctuations. Nevertheless, we demonstrate in this paper that 
the excellent achieved sensitivity makes it possible to perform reconstruction of 
the lensing potential using only information at larger angular scales, and report a significant detection
in the auto-spectrum of the reconstructed lensing potential.
In addition, we cross-correlate our reconstructed lensing map with the published
\planck\ lensing potential \citep{P15:phi}. This cross-spectrum, which is immune to most systematic effects and foregrounds, 
also detects lensing with high significance. Since the \planck\ lensing potential is reconstructed primarily 
using temperature, and that from \BK\ is reconstructed entirely using polarization, 
the strong correlation of the two maps shows that they are producing a consistent reconstruction of the true lensing potential.
The derived lensing amplitudes are consistent with that expected in the $\Lambda$CDM cosmology. 
Taken together, these results imply that the $B$-mode power in the multipole range of $150\lesssim\l\lesssim350$
previously detected by \BK\ \citep{BKVI} is indeed caused by lensing. 

This paper is part of an on-going series describing results and methods from the \BK\ series of experiments 
(\citealt{B2I}, hereafter BK-I; \citealt{B2II}, hereafter BK-II; \citealt{B2III}, hereafter BK-III; 
\citealt{BKIV}, hereafter BK-IV; \citealt{BKV}, hereafter BK-V; \citealt{BKP}, hereafter BKP;
\citealt{BKVI}, hereafter BK-VI; \citealt{BKVII}, hereafter BK-VII).
This paper is organized as follows:
in \sec{data} we briefly summarize the data sets that are used in this paper, 
in \sec{analysis} we describe our analysis method for reconstructing the lensing potential from the \BK\ data,
in \sec{results} we give our results including the auto- and cross-spectra of the lensing potential,
in \sec{systematics} we present consistency and null tests,
and in \sec{conclusions} we conclude.

%////////////////////////////////////////////////////////////////////////////////////////////////////%
\section{Observed data and simulations} \label{data}
%////////////////////////////////////////////////////////////////////////////////////////////////////%

%::::::::::::::::::::::::::::::::::::::::::::::::::::::::::::::::::::::::::::::::::::::::::::::::::::%
\subsection{\biceptwo\ and \keckarray}
%::::::::::::::::::::::::::::::::::::::::::::::::::::::::::::::::::::::::::::::::::::::::::::::::::::%

In this paper we use the \BK\ maps which coadd all data taken up to and including
the 2014 observing season---we refer to these as the \bk\ maps.
These maps were previously described in BK-VI,
where they were converted to power spectra, and used to set constraints
on the amplitudes of primordial $B$-modes and foregrounds.
In this work we use only the 150\,GHz $Q/U$ maps.
These have a depth of $3.0\ \mu$K-arcmin over an effective area of $\sim 395$ deg$^2$, 
centered on RA 0h, Dec.\ $-57.5^{\circ}$. 

We re-use the standard sets of simulations described in BK-VI and previous papers:
lensed and unlensed CMB signal-only simulations (denoted by ``lensed/unlensed-$\Lambda$CDM''),
instrumental noise, and dust foreground, each having $499$ realizations.
In addition in this paper we also make use of the input lensing potential. 
The details of the signal and noise simulations are given in Sec.~V of BK-I 
and the dust simulations are described in Sec.~IV.A of BKP
and Appendix E of BK-VI. As discussed in \sec{analysis}, 
the lensed-$\Lambda$CDM, instrumental noise, and dust simulated maps are combined  
to estimate the transfer function, mean-field bias, disconnected bias, 
and the uncertainties of the lensing power spectrum.
The unlensed-$\Lambda$CDM simulations are used to evaluate the
significance of detection of lensing (rejection of the no-lensing hypothesis).
Lensing is applied to the unlensed input maps using Lenspix~\citep{Lewis:2005}
as described in Sec.~V.A.2 of BK-I.

Starting with the spherical harmonic coefficients of the input lensing potential
(from Lenspix) we first transform to the lensing-mass field $\kappa$ (lensing convergence) using 
%----------------------------------------------------------------------------------------------------%
\al{
	\kappa_{LM} = -\frac{L(L+1)}{2}\grad_{LM} \,, \label{eq:kappa}
}
%----------------------------------------------------------------------------------------------------%
and then make the full-sky $\kappa$ map by the spherical harmonic transform of $\kappa_{LM}$. 
This transformation is necessary to avoid mode mixing in the subsequent apodization to the
\bk\ sky patch because the lensing-mass field has a nearly flat spectrum, while the lensing potential
has a red spectrum \citep{P15:phi}. 
Next the input lensing-mass map in the \bk\ sky patch, $\kappa^{\rm in}(\hatn)$, 
is obtained by interpolating the full-sky $\kappa$ map to the standard
\bk\ map pixelization, and multiplying by the standard inverse variance apodization mask.
Here $\hatn$ denotes position in the \bk\ sky patch. 
%Note that the difference between the power spectra of the full sky and projected lensing-mass map
%could be generated by the apodization window but is found to be very small. 
Finally, the Fourier modes of the input lensing potential in the \bk\ sky patch, $\grad^{\rm in}_{\bL}$, are 
calculated from 
%----------------------------------------------------------------------------------------------------%
\al{
	\grad^{\rm in}_{\bL} = -\frac{2}{L^2}\FT{\hatn}{\bL}{f} \kappa^{\rm in}(\hatn) \,. \label{Eq:massmap}
}
%----------------------------------------------------------------------------------------------------%
Here and after, we use $L$ for the multipoles of the lensing potential and $\l$ for the $E$ and $B$ modes. 

%::::::::::::::::::::::::::::::::::::::::::::::::::::::::::::::::::::::::::::::::::::::::::::::::::::%
\subsection{\planck}
%::::::::::::::::::::::::::::::::::::::::::::::::::::::::::::::::::::::::::::::::::::::::::::::::::::%

We use the publicly available \planck\ 2015 lensing-mass field \citep{P15:phi}.
This lensing-mass field is estimated by optimally combining 
all of the quadratic estimators constructed from the SMICA temperature and $E$/$B$ maps. 
The most effective of the estimators is $TT$, 
but the $TE$ and $EE$ estimators also improve the total significance of the detection. 
We also use the \planck\ 2013 lensing potential \citep{P13:phi},
which has larger statistical uncertainty, but, since it is reconstructed using
the temperature maps only, is a useful cross check.

The publicly released \planck\ 2015 lensing package contains multipole coefficients for the observed 
lensing-mass field as well as $100$ simulated realizations of input and reconstructed lensing-mass fields.
The \planck\ 2013 release instead provides multipole coefficients of the unnormalized lensing potential,
so we multiply by the provided response function (see Sec.~2 of \citealt{P13:phi}), 
and make a full sky lensing-mass field.
The full-sky \planck\ lensing-mass maps, with point sources masked, are interpolated to
the standard \bk\ map pixelization.
We find that the noise contribution to the \planck\ lensing-mass map in this region 
is approximately $\sim 20$\% smaller than that of the full-sky average
due to the scan strategy of the \planck\ mission. 

As discussed in \sec{analysis}, the \planck\ simulations are used to evaluate 
the expected correlation between the \bk\ and \planck\ lensing signals and its statistical uncertainty. 
In order to correlate the reconstructed lensing signal between the \bk\ and \planck\ simulations, 
we replace each \planck\ lensing realization with those of the \bk\ simulations using
(e.g., \citealt{DESxPlanck,Kirk:2015}) 
%----------------------------------------------------------------------------------------------------%
\al{
	\widehat{\kappa}^{\rm sim,P'}(\hatn) = \widehat{\kappa}^{\rm sim,P}(\hatn) - \kappa^{\rm in,P}(\hatn) + \kappa^{\rm in}(\hatn) \,,
}
%----------------------------------------------------------------------------------------------------%
where $\kappa^{\rm in,P}$ and $\widehat{\kappa}^{\rm sim,P}$ are the input and reconstructed lensing-mass maps of 
the \planck\ simulations, and $\kappa^{\rm in}$ is the input lensing-mass map of the \bk\ realizations.
We checked that the correlation between $\widehat{\kappa}^{\rm sim,P'}$ and $\kappa^{\rm in,P}$ is consistent with zero. 
We then multiply $\widehat{\kappa}^{\rm sim,P'}$ by the standard \bk\ inverse variance apodization mask,
and Fourier transform according to \eq{Eq:massmap}.

Hereafter, unless otherwise stated, the \planck\ data refers to the \planck\ 2015 release products.

%////////////////////////////////////////////////////////////////////////////////////////////////////%
\section{Lensing reconstruction method} \label{analysis}
%////////////////////////////////////////////////////////////////////////////////////////////////////%

It is possible to reconstruct the lensing potential from observed CMB anisotropies because
lensing introduces off-diagonal mode-mode covariance within, and between, the
$T$- $E$- and $B$- mode sets.
An estimator of the lensing potential is then given by a quadratic form in the CMB anisotropies.
The power spectrum of the lensing potential $C_L^{\grad\grad}$ (lensing potential power spectrum)
can be studied by taking the power spectrum of the lensing potential estimator. 

In this section we describe the method used to reconstruct the lensing potential
from the \bk\ polarization map, to calculate the lensing potential power spectrum, and to
evaluate the amplitudes of the resulting power spectra for the data sets described in \sec{data}.

%::::::::::::::::::::::::::::::::::::::::::::::::::::::::::::::::::::::::::::::::::::::::::::::::::::%
\subsection{Lensed CMB anisotropies}
%::::::::::::::::::::::::::::::::::::::::::::::::::::::::::::::::::::::::::::::::::::::::::::::::::::%

The effect of lensing on the $Q$ and $U$ maps is given by (e.g., \citealt{Lewis:2006fu,Hanson:review})
%----------------------------------------------------------------------------------------------------%
\al{
	[\widetilde{Q}\pm\iu \widetilde{U}](\hatn) &= [Q\pm\iu U](\hatn+\bd(\hatn)) \notag \\
		&\simeq [Q\pm\iu U](\hatn) + \bd(\hatn)\cdot\bn [Q\pm\iu U](\hatn)
	\,, \label{eq:remap}
}
%----------------------------------------------------------------------------------------------------%
where $\hatn$ is the observed direction, and $Q$($\widetilde{Q}$) and $U$($\widetilde{U}$) are 
the unlensed (lensed) anisotropies. The two-dimensional vector $\bd(\hatn)$ is 
the deflection angle, with two degrees of freedom. In terms of parity symmetry, these two components 
are given as the lensing potential (even parity), and curl-mode deflection (odd parity) \citep{Hirata:2003ka}: 
%----------------------------------------------------------------------------------------------------%
\al{
	\bn^2 \grad(\hatn) &= \bn\cdot\bd(\hatn)  \,, \label{eq:grad} \\
	(\star\bn)^2 \curl(\hatn) &= (\star\bn)\cdot\bd(\hatn)  \,, \label{eq:curl}
}
%----------------------------------------------------------------------------------------------------%
where $\bn$ is the covariant derivative on the sphere, and $\star$ denotes the operation
that rotates the angle of a two-dimensional vector counterclockwise by $90$ degrees.

The $E$ and $B$ modes are defined as 
%----------------------------------------------------------------------------------------------------%
\al{
	E_{\bl} \pm i B_{\bl} = - \FT{\hatn}{\bl}{f} [Q\pm\iu U](\hatn) \E^{\mp 2\iu\varphi_{\bl}}  \,,
}
%----------------------------------------------------------------------------------------------------%
where $\varphi_{\bl}$ is the angle of $\bl$ measured from the Stokes $Q$ axis.
With the lensing potential and curl mode given in Eqs.~\eqref{eq:grad} and \eqref{eq:curl},
the lensed $E$ and $B$ modes are given by (e.g., \citealt{Hu:2001kj,Cooray:2005hm})
%----------------------------------------------------------------------------------------------------%
\al{
	\tE_{\bl} &= E_{\bl} + \Int{2}{\bL}{(2\pi)^2}
		[\bL\cdot(\bl-\bL)\grad_{\bL} + (\star\bL)\cdot(\bl-\bL)\curl_{\bL}]
	\notag \\
		&\qquad \times E_{\bl-\bL}\cos 2(\varphi_{\bl-\bL}-\varphi_{\bl})
	\label{Eq:lensing-E} \\
	\tB_{\bl} &= \Int{2}{\bL}{(2\pi)^2}
		[\bL\cdot(\bl-\bL)\grad_{\bL} + (\star\bL)\cdot(\bl-\bL)\curl_{\bL}]
	\notag \\
		&\qquad \times E_{\bl-\bL}\sin 2(\varphi_{\bl-\bL}-\varphi_{\bl})
	\,. \label{Eq:lensing-B}
}
%----------------------------------------------------------------------------------------------------%
Because the contribution of $B$-modes from gravitational waves is tightly constrained in the BK-VI paper,
and rapidly decreases in amplitude at $\l>100$, we ignore their possible contribution here.

Up to first order in $\grad$ and $\curl$, the lensing-induced off-diagonal elements of the covariance 
are (e.g., \citealt{Hu:2001kj,Cooray:2005hm})
%----------------------------------------------------------------------------------------------------%
\al{
	\ave{\tE_{\bL}\tB_{\bl-\bL}}_{\rm CMB} &= w_{\bl,\bL}^\grad \grad_{\bl} + w_{\bl,\bL}^\curl \curl_{\bl}
	\,, \label{Eq:weight}
}
%----------------------------------------------------------------------------------------------------%
where $\ave{\cdots}\rom{CMB}$ denotes the ensemble average over unlensed $E$-modes, 
with a fixed realization of the lensing potential and curl modes.
The explicit forms of the weight functions for the lensing potential and curl mode 
are, respectively, given in \citet{Hu:2001kj} and \citet{Namikawa:2011cs} as
%----------------------------------------------------------------------------------------------------%
\al{
	w^\grad_{\bL,\bl} &= -\bl\cdot(\bL-\bl)\tCEE_\l\sin 2(\varphi_{\bl}-\varphi_{\bL-\bl})
	\,, \label{Eq:weight:g} \\
	w^\curl_{\bL,\bl} &= -(\star\bl)\cdot(\bL-\bl)\tCEE_\l\sin 2(\varphi_{\bl}-\varphi_{\bL-\bl})
	\,, \label{Eq:weight:c}
}
%----------------------------------------------------------------------------------------------------%
where $\tCEE_\l$ is the lensed $E$-mode power spectrum to take into account the higher order biases \citep{Hanson:2010rp,Lewis:2011}. 
\eq{Eq:weight} means that the lensing signals, $\grad$ and $\curl$, can be estimated through off-diagonal elements of 
the covariance matrix of the CMB Fourier modes (see \sec{Sec:est} for details).
Note that we do not include $\curl$ in our simulations because its contribution is negligible 
in the standard $\Lambda$CDM model (e.g. \citealt{Saga:2015,Pratten:2016}). 
We use the reconstructed curl mode as a null test in \sec{systematics}. 

%::::::::::::::::::::::::::::::::::::::::::::::::::::::::::::::::::::::::::::::::::::::::::::::::::::%
\subsection{Input E and B-modes for reconstruction} 
%::::::::::::::::::::::::::::::::::::::::::::::::::::::::::::::::::::::::::::::::::::::::::::::::::::%

In \BK\ analysis, we use real space matrix operations to process the data into purified $E$- and $B$-maps, 
which are then transformed to multipole space. The sky signal is filtered by the observing strategy, 
and the analysis process, including the removal of potential systematic errors (``deprojection"). 
These effects are entirely captured in an {\it observation matrix}, $\bR{\mathcal{R}}$ 
(see \citealt{Tolan:PhD} and BK-VII). 
The observed maps, $Q^{\rm obs}$ and $U^{\rm obs}$, are then given by
%----------------------------------------------------------------------------------------------------%
\al{
	\binom{Q^{\rm obs}}{U^{\rm obs}} = \bR{\mathcal{R}}\binom{Q'}{U'} + \binom{Q^{\rm noise}}{U^{\rm noise}}  \,. 
}
%----------------------------------------------------------------------------------------------------%
Here $Q'$ and $U'$ are an input signal realization---in this case lensed-$\Lambda$CDM+dust---and the
second term is a noise realization.
The observed map suffers from some mixing of $E$- and $B$-modes induced by 
e.g. the survey boundary and the filtering.
To mitigate the mixing between $E$- and $B$-modes, 
the observed $E$- and $B$-mode maps are multiplied by purification matrices $\bR{\Pi}^{\rm E}$ 
and $\bR{\Pi}^{\rm B}$ respectively to recover {\it pure} $E$- and $B$-modes.
This operation is simply expressed as \citep{Tolan:PhD}
%----------------------------------------------------------------------------------------------------%
\al{
	\binom{\widehat{Q}^{\rm E}}{\widehat{U}^{\rm E}} = \bR{\Pi}^{\rm E}\binom{Q^{\rm obs}}{U^{\rm obs}} 
	\,, \label{Eq:purifQ} \\
	\binom{\widehat{Q}^{\rm B}}{\widehat{U}^{\rm B}} = \bR{\Pi}^{\rm B}\binom{Q^{\rm obs}}{U^{\rm obs}} 
	\,, \label{Eq:purifU}
}
%----------------------------------------------------------------------------------------------------%
where $\widehat{Q}^{\rm E}$ ($\widehat{Q}^{\rm B}$) and $\widehat{U}^{\rm E}$ ($\widehat{U}^{\rm B}$)
are purified Stokes $Q$ and $U$ maps containing as much of the original $E$ ($B$) modes as possible. 
The purified $Q$ and $U$ maps are further multiplied by the standard inverse variance apodization mask
to downweight noisy pixels around the survey boundary.
The Fourier transforms of the purified, apodized $Q$/$U$ maps are converted to purified $E$- and $B$-modes, 
$\hE$ and $\hB$, and these are used as inputs to the lensing reconstruction analysis.

The input CMB Fourier modes require proper weighting to optimize the lensing reconstruction. 
In the ideal case (i.e., white noise, full-sky observation with no filtering), 
the lensing reconstruction is optimized by a simple diagonal weighting of $E_{\bl}$ and $B_{\bl}$. 
Denoting $X=E$ or $B$, the optimally-weighted Fourier modes are given by \citep{Hu:2001kj}
%----------------------------------------------------------------------------------------------------%
\al{
	\bX_{\bl} &= \frac{\hX_{\bl}}{\hCXX_\l}  \,. 
}
%----------------------------------------------------------------------------------------------------%
Here $\hE$ and $\hB$ are, again, the purified $E$ and $B$-modes obtained by 
the Fourier transform of the purified $Q$/$U$ maps in Eqs.~\eqref{Eq:purifQ} and \eqref{Eq:purifU}, 
and $\hCXX_\l$ is an isotropic power spectrum including noise and foregrounds. 
In more realistic situations, using diagonal filtering degrades the sensitivity to 
the lensing potential \citep{Hirata:2003ka,Smith07,Hanson:2009gu,P13:phi}. 
To take into account the anisotropic filtering and noise 
we multiply two-dimensional functions in Fourier space to the purified $E$- and $B$-modes 
%----------------------------------------------------------------------------------------------------%
\al{
	\bX_{\bl} &\simeq \frac{f^{\rm X}_{\bl}\hX_{\bl}}{\hCXX_{\bl}} \,, \label{Eq:bX}
}
%----------------------------------------------------------------------------------------------------%
where $\hCXX_{\bl}$ is the mean of the two dimensional $E$-/$B$-mode spectra 
of the lensed-$\Lambda$CDM+dust+noise simulations, and the factor $f^{\rm X}_{\bl}$ describes 
the beam and filtering suppression of the $E$- and $B$-modes. 
We calculate these suppression factors by comparing the mean input and output power spectra of 
the lensed-$\Lambda$CDM signal-only simulations, 
$C^{\rm XX,in}_\l$ and $C^{\rm XX,out}_{\bl}$, as $(f^{\rm X}_{\bl})^2 = C^{\rm XX,out}_{\bl}/C^{\rm XX,in}_\l$. 

%<><><><><><><><><><><><><><><><><><><><><><><><><><><><><><><><><><><><><><><><><><><><><><><><><><>%
\begin{figure} 
\bc
\includegraphics[width=8.5cm,clip]{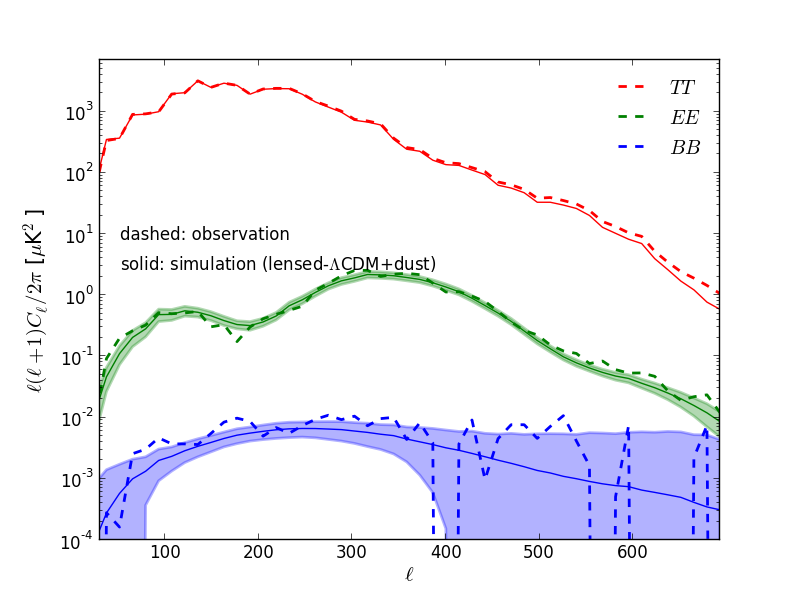}
\caption{
Raw temperature, $E$- and $B$-mode spectra (dashed lines) measured from \bk, 
compared with those from lensed-$\Lambda$CDM+dust+noise simulations.
(Both have been noise debiased but no beam or filtering correction has been applied.)
For the simulations the solid line shows the mean value and the shaded region the $\pm 1\sigma$ range.
The temperature sky input to the simulations is constrained to be the actual sky pattern
as observed by \planck\ so there is no sample variance in the simulated $TT$ spectra.
}
\label{Fig:EE:BB}
\ec
\end{figure}
%<><><><><><><><><><><><><><><><><><><><><><><><><><><><><><><><><><><><><><><><><><><><><><><><><><>%

In addition to the above filtering function, we filter in multipole space to select $E$ and $B$-modes
with the baseline ranges being $30\leq\l\leq700$ and $150\leq\l\leq700$ respectively. 
The minimum multipole of the $E$-modes is set by the timestream filtering---multipoles smaller
than $\lsim 30$ are so heavily attenuated as to be unrecoverable.
The minimum multipole of the $B$-modes is chosen so that the dust foreground is 
subdominant compared to the lensing $B$-modes. 
The nominal maximum multipole is set by the resolution of the standard \bk\ maps which have 0.25$^\circ$ pixel spacing.
We will see later (in \sec{systematics}) that restricting to $\l<600$ makes very little
difference to the final result.

We have not previously published any results for $\l>350$ because the beam correction becomes very large, 
and hence in principle so does the uncertainty on that correction.
As shown in \fig{Fig:EE:BB}, we find that the mean of the signal simulations actually
remains very close to the observed bandpower values for multipoles all the way up to the pixel scale.
However there is a small positive deviation at higher $\l$ which reaches $20\%$ at $\l=600$
implying that we have slightly under-estimated our beam function in this range.
This is very clear in the $TT$ spectrum because the input sky for the simulations
is constrained to the actual sky pattern as observed by \planck\ (as described in Sec.~V.A.1 of BK-I),
and hence there is no sample variance in this comparison.
Based on this observation we apply a small additional beam correction
for the baseline lensing analysis presented in this paper.
In practice, we multiply the inverse square root 
of the $\l$-dependent correction to the observed (and also simulated noise) $E$-/$B$-modes, and then 
compute the weighted Fourier modes of \eq{Eq:bX}.
As shown in \sec{systematics}, this correction only leads to small changes 
in the final results.

%::::::::::::::::::::::::::::::::::::::::::::::::::::::::::::::::::::::::::::::::::::::::::::::::::::%
\subsection{Estimating the lensing potential} \label{Sec:est}
%::::::::::::::::::::::::::::::::::::::::::::::::::::::::::::::::::::::::::::::::::::::::::::::::::::%

We now describe the estimator for the lensing potential. 
Eq.~\eqref{Eq:weight} motivates the following quadratic estimator for the lensing potential \citep{Hu:2001,Hu:2001kj} 
%----------------------------------------------------------------------------------------------------%
\al{
	\estg_{\bL} = A_{\bL}^\grad (\uestg_{\bL}-\ave{\uestg_{\bL}})  \,, \label{Eq:est}
}
%----------------------------------------------------------------------------------------------------%
where $\ave{\cdots}$ is the ensemble average over realizations of purified $E$ and $B$ modes, 
and $\uestg_{\bL}$ is the unnormalized $EB$ estimator
%----------------------------------------------------------------------------------------------------%
\al{
	\uestg_{\bL} &= \Int{2}{\bl}{(2\pi)^2} w^\grad_{\bL,\bl}\bE_{\bl}\bB_{\bL-\bl} 
	\,. \label{Eq:def-AXY}
}
%----------------------------------------------------------------------------------------------------%
Here $w^\grad_{\bL,\bl}$ is the weight function given in \eq{Eq:weight:g}. 
The second term, $\ave{\uestg_{\bL}}$, is a correction for the mean-field bias, and is estimated from the simulations. 
The quantities, $\bE$ and $\bB$, are the weighted Fourier modes given in \eq{Eq:bX}, 
and $A_{\bL}$ is a normalization that makes the estimator unbiased. 

Similarly, the curl-mode estimator is constructed by replacing the weight function with 
$w_{\bL,\bl}^\curl$, which is given in \eq{Eq:weight:c}. 
Up to first order in $\grad$ and $\curl$, the estimator of the lensing potential
is unbiased even in the presence of the curl-mode, and vice versa \citep{Namikawa:2011cs}.

Unlike the lensing reconstruction from the temperature and $E$-mode, 
the mean-field bias due to the presence of the sky cut is typically small for this $EB$ estimator 
with an appropriate treatment for $E$/$B$ mixing \citep{Namikawa:2013,Pearson:2014}. 
Other non-lensing anisotropies could generate a mean-field component (e.g. \citealt{Hanson:2009}), 
but our simulations show that the mean-field bias is smaller than the simulation noise 
which corresponds to $A_L^\grad$ divided by the number of realizations (see e.g. \citealt{Namikawa:2012pe}). 
We also note again that our simulated maps are generated with the temperature sky constrained to
that observed by \planck.
However, the use of these constrained realizations results in a contribution in the mean-field bias 
which is consistent with the simulation noise, and therefore has a negligible effect on our results.

In the ideal case, the normalization of the estimator is given analytically by 
%----------------------------------------------------------------------------------------------------%
\al{
	\ol{A}_{\bL}^\grad &= \left\{\Int{2}{\bl}{(2\pi)^2} \frac{|w_{\bL,\bl}^{\grad}|^2}
		{\hCEE_\l\hCBB_{|\bL-\bl|}}\right\}^{-1}  \,. \label{Eq:norm-theory}
}
%----------------------------------------------------------------------------------------------------%
In \BK, different CMB multipoles are mixed by the survey boundary and anisotropic filtering. 
Therefore, we calculate the normalization factor using simulations, as other 
experiments have done (\citealt{PB14:phi,vanEngelen:2014zlh,Story:2014hni}). In practice, 
we use the following additional normalization:
%----------------------------------------------------------------------------------------------------%
\al{
	A_{\bL}^\grad = \frac{\ave{|\grad^{\rm in}_{\bL}|^2}}{\ave{\grad^{\rm in}_{\bL}(\estg_{\bL}^{\rm sim})^*}} \ol{A}^\grad_{\bL}
	\,, \label{Eq:transfer}
}
%----------------------------------------------------------------------------------------------------%
where $\grad_{\bL}^{\rm in}$ and $\estg_{\bL}^{\rm sim}$ are the input and reconstructed lensing potential from simulation. 

%::::::::::::::::::::::::::::::::::::::::::::::::::::::::::::::::::::::::::::::::::::::::::::::::::::%
\subsection{Estimating the lensing potential power spectrum} 
%::::::::::::::::::::::::::::::::::::::::::::::::::::::::::::::::::::::::::::::::::::::::::::::::::::%

We estimate the lensing potential power spectrum using the reconstructed lensing potential from \bk\ data alone, 
and also by cross-correlating the reconstructed lensing potential from \bk\ with that from \planck. 

The power spectrum of the lensing potential is estimated by squaring $\estg_{\bL}$. 
The lensing potential estimator is quadratic in the CMB, and its power spectrum is the four-point correlation of the 
CMB anisotropies. This power spectrum can be decomposed into the disconnected and connected parts 
%----------------------------------------------------------------------------------------------------%
\al{
	\ave{|\estg_{\bL}|^2} = \ave{|\estg_{\bL}|^2}_{\rm C} + \ave{|\estg_{\bL}|^2}_{\rm D} 
	\,. \label{Eq:estg-power}
}
%----------------------------------------------------------------------------------------------------%
The disconnected part $|\estg_{\bL}|^2_{\rm D}$ comes from the Gaussian part of the four-point 
correlation, while the connected part contains the non-Gaussian contributions from lensing. 
The connected part gives the lensing power spectrum, $C_L^{\grad\grad}$, 
with a correction from the higher-order bias \citep{Kesden:2003} which is negligible in our analysis. 
On the other hand, the disconnected part of the four-point correlation remains 
even in the absence of lensing, and is given by 
%----------------------------------------------------------------------------------------------------%
\al{
	\ave{|\estg|^2_{\bL}}_{\rm D} &= (A_L^\grad)^2
		\Int{2}{\bl}{(2\pi)^2} \Int{2}{\bl'}{(2\pi)^2} w^{\grad}_{\bL,\bl}w^{\grad}_{-\bL,\bl'}
	\notag \\
	&\times 
		[\ol{\bR{C}}^{\rm EE}_{\bl,\bl'}\ol{\bR{C}}^{\rm BB}_{\bL-\bl,-\bL-\bl'}
		+\ol{\bR{C}}^{\rm EB}_{\bl,\bL-\bl'}\ol{\bR{C}}^{\rm EB}_{\bl',-\bL-\bl}]
	\,, \label{Eq:disconnect}
}
%----------------------------------------------------------------------------------------------------%
where $A_L^\grad$ is the angle average of the estimator normalization in \eq{Eq:transfer}, and 
$\ol{\bR{C}}^{XY}_{\bl,\bl'}\equiv\ave{\bX_{\bl}\bY_{\bl'}}$ is the covariance matrix. 
The disconnected part of the four-point correlation is produced by both the CMB fluctuations 
and instrumental noise. 
We describe the treatment of the disconnected bias for auto- and cross- lensing power 
in the next two sections.  

%++++++++++++++++++++++++++++++++++++++++++++++++++++++++++++++++++++++++++++++++++++++++++++++++++++%
\subsubsection{Auto-spectrum of \bk}
%++++++++++++++++++++++++++++++++++++++++++++++++++++++++++++++++++++++++++++++++++++++++++++++++++++%

For the auto-spectrum of the \bk\ lensing potential, the disconnected bias is a significant 
contribution that must be subtracted.
The de-biased lensing potential power spectrum is given by 
%----------------------------------------------------------------------------------------------------%
\al{
	\hckk_{\bL} \equiv |\estg_{\bL}|^2-\hN_{\bL}^\grad  \,, \label{Eq:est-clgg}
}
%----------------------------------------------------------------------------------------------------%
where $\hN_{\bL}^{\grad}$ is the disconnected bias, and a normalization factor 
(the correction for the apodization window) is omitted for clarity.
In principle, this Gaussian bias can be estimated from the explicit formula in \eq{Eq:disconnect}
or dedicated Gaussian simulations. 
However, these approaches rely on an accurate model of $\ol{\bR{C}}^{XY}_{\bl,\bl'}$. 
Using an inaccurate covariance matrix, $\ol{\bR{C}}^{XY}+\Sigma^{XY}$, \eq{Eq:disconnect}
results in an error $\mC{O}(\Sigma^{XY})$. 

In our analysis, the disconnected bias is estimated with the {\it realization-dependent} method 
developed by \citet{Namikawa:2012pe} for temperature and extended by \citet{Namikawa:2013}
to include polarization. 
In this method, part of the covariance is replaced with the real data, and is given by
%----------------------------------------------------------------------------------------------------%
\al{
	\hN_{\bL}^\grad &= \ave{|\estg^{E_{\bm{1}},\hB}_{\bL}+\estg_{\bL}^{\hE,B_{\bm{1}}}|^2}_{\bm{1}} 
		- \frac{1}{2}\ave{|\estg_{\bL}^{E_{\bm{1}},B_{\bm{2}}}+\estg_{\bL}^{E_{\bm{2}},B_{\bm{1}}}|^2}_{\bm{1},\bm{2}} 
	\,. \label{Eq:hN}
}
%----------------------------------------------------------------------------------------------------%
Here $\estg^{XY}_{\bL}$ is the lensing estimator computed from the quadratic combination of $X$ and $Y$. 
$\hE$ and $\hB$ are the purified $E$- and $B$-modes from real data, 
while $E_{\bm 1}$ ($E_{\bm 2}$) and $B_{\bm 1}$ ($B_{\bm 2}$) 
are generated from the first (second) set of simulations. 
The ensemble average $\ave{\cdots}_{\bm i}$ is taken over the $i$'th set of simulations. 
Our simulation set is divided into two subsets multiple times to estimate the second term. 

Note that this form of disconnected bias is obtained naturally from the optimal estimator for 
the lensing-induced trispectrum using the Edgeworth expansion of the CMB likelihood (Appendix A). 
Realization-dependent methods have the benefit of suppressing spurious off-diagonal elements 
in the covariance matrix. 
Furthermore, the disconnected bias estimated using this method is less sensitive to the accuracy of 
the covariance, i.e., it contains contributions from $\mC{O}(\Sigma^2)$ instead of $\mC{O}(\Sigma)$.

The curl-mode power spectrum is also estimated in the same way but with the quadratic estimator of 
the curl-mode $\estc_{\bL}$, while the disconnected bias becomes very small 
in estimating the cross-spectrum between the lensing potential and curl mode. 

%++++++++++++++++++++++++++++++++++++++++++++++++++++++++++++++++++++++++++++++++++++++++++++++++++++%
\subsubsection{Cross-spectrum with \planck}
%++++++++++++++++++++++++++++++++++++++++++++++++++++++++++++++++++++++++++++++++++++++++++++++++++++%

In cross-correlation studies involving \planck\, we expect the disconnected bias in
cross-spectra to be completely negligible. The reasons are as follows.

In cross-spectrum analysis, the instrumental noise of the two experiments is uncorrelated.
Disconnected bias can only arise from sky signal. The \planck\ 2015 lensing potential is estimated
from all of the quadratic estimators, including those involving polarization.
Therefore, even in the absence of lensing, two of these quadratic estimators (the $EB$ and $TB$ estimators)
are correlated with the $EB$ estimator computed from the \bk\ data through the common sky signal.

In practice, this disconnected bias is small. The correlation of $B$-modes between these two experiments 
does not contain noise contributions. The four-point correlation, $EBEB$ and $TBEB$, are then produced by
the CMB $B$-mode signals but not by the instrumental noise in $B$-modes.
The uncertainties in the \planck\ lensing potential are dominated by instrumental noise, which is much larger
than any possible $B$-modes on the sky that can lead to a disconnected bias.
To see this more quantitatively, we evaluate the disconnected
bias expected from the $\Lambda$CDM $B$-mode power spectrum and appropriate noise levels, 
using the analytic formula based on \citet{Hu:2001kj}.
We find that the bias is indeed negligible compared to the reconstruction noise (see \fig{Fig:planck}).

In addition, since the \planck\ 2013 lensing potential is reconstructed from the temperature maps alone,
the cross-spectrum between \bk\ and \planck\ 2013 is free of any disconnected bias.
In the next section, we show that the cross-spectrum results with \planck\ 2013 and \planck\ 2015 are consistent,
again confirming that the disconnected bias in the \planck\ 2015 - \bk\ cross-spectrum is not
significant.

%Since unlensed CMB fluctuations in our simulation are not correlated with that in 
%the \planck\ simulation, the N1 bias \citep{Kesden:2003} is not included in the simulated cross-power spectrum. 
%This bias is, however, negligible in our results due to the following reasons.
%As shown in \citep{Jenkins:2014}, 
%the N1 bias from a four-point correlation which has one $B$-mode (such as $TTEB$) is several orders of magnitudes 
%smaller than the lensing signal power spectrum. 
%In our analysis, the N1 bias in the cross-power spectrum 
%between \bk and \planck\ 2013 comes from $TTEB$, and is therefore negligible. 
%On the other hand, the cross-power spectrum with \planck\ 2015 has the N1 bias from $EBEB$ and $TBEB$
%but these contributions are significantly suppressed in the minimum variance estimator. 

%++++++++++++++++++++++++++++++++++++++++++++++++++++++++++++++++++++++++++++++++++++++++++++++++++++%
\subsubsection{Binned power spectrum and its amplitude}
%++++++++++++++++++++++++++++++++++++++++++++++++++++++++++++++++++++++++++++++++++++++++++++++++++++%

In our analysis the multipoles between $30$ and $700$ are divided into $10$ bins and 
the bandpowers of the lensing potential power spectrum $C_b$ are given at these multipole bins. 
We estimate the amplitude of the lensing potential power spectrum as a weighted mean over multipole bins
%----------------------------------------------------------------------------------------------------%
\al{
	\AL = \frac{\sum_b a_b A_b}{\sum_b a_b}  \,, \label{Eq:AL}
}
%----------------------------------------------------------------------------------------------------%
where $A_b$ is the relative amplitude of the power spectrum compared with a fiducial power 
spectrum $C_b^{\rm f}$, i.e., $A_b \equiv C_b/C_b^{\rm f}$, and the weights, $a_b$, are taken
from the bandpower covariance according to
%----------------------------------------------------------------------------------------------------%
\al{
	a_b = \sum_{b'} C^{\rm f}_b \bR{Cov}_{bb'}^{-1} C^{\rm f}_{b'}  \,.
}
%----------------------------------------------------------------------------------------------------%
The fiducial bandpower values and their covariances are evaluated from the simulations.
Consequently, $\AL$ defined as above is an amplitude relative to 
the \planck\ $\Lambda$CDM prediction.

%////////////////////////////////////////////////////////////////////////////////////////////////////%
\section{Results} \label{results}
%////////////////////////////////////////////////////////////////////////////////////////////////////%

%<><><><><><><><><><><><><><><><><><><><><><><><><><><><><><><><><><><><><><><><><><><><><><><><><><>%
\begin{figure} 
\bc
\includegraphics[width=8.5cm,clip]{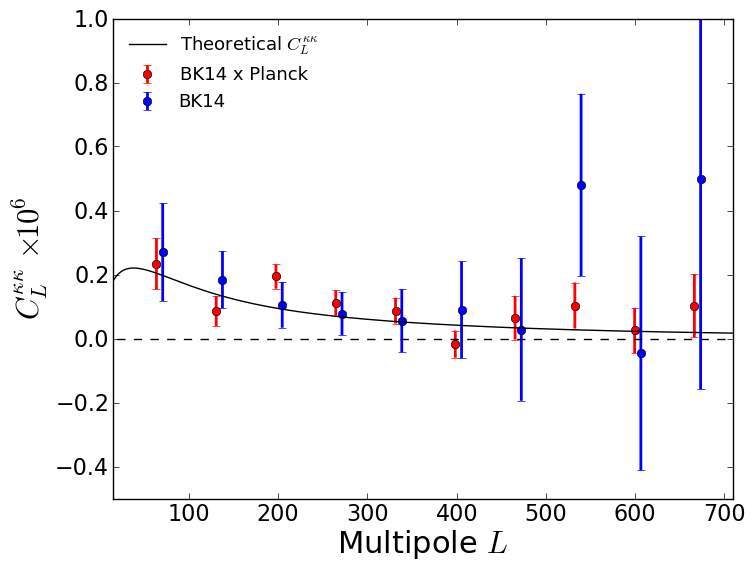}
\caption{
The lensing-mass power spectrum, $C_L^{\kappa\kappa}=L^4C^{\grad\grad}_L/4$, 
estimated from the cross-spectrum between \bk\ and \planck\ 2015 data (red), 
and the auto-spectrum of \bk\ data (blue). 
The black solid line shows the theoretical spectrum assuming the $\Lambda$CDM cosmology.
The \bk\ auto-spectrum is offset in $L$ for clarity. 
}
\label{Fig:ClKK}
\ec
\end{figure}
%<><><><><><><><><><><><><><><><><><><><><><><><><><><><><><><><><><><><><><><><><><><><><><><><><><>%

%<><><><><><><><><><><><><><><><><><><><><><><><><><><><><><><><><><><><><><><><><><><><><><><><><><>%
\begin{table}
\renewcommand{\arraystretch}{1.8}
\bc
\caption{
The bandpowers of the lensing-mass power spectrum and $1\sigma$ statistical errors 
at the center of each bin, $L_{\rm c}$, as shown in \fig{Fig:ClKK}. 
The values of the bandpowers and errors are multiplied by $10^7$. 
}
\label{table:clkk}
\begin{tabular}{lcc} \hline 
$L_{\rm c}$ & \bk$\times$\planck\ & \bk \\ \hline 
$63.5$  & $2.33\pm0.80$  & $2.70\pm1.53$ \\
$130.5$ & $0.86\pm0.47$  & $1.84\pm0.89$ \\
$197.5$ & $1.94\pm0.38$  & $1.05\pm0.72$ \\
$264.5$ & $1.11\pm0.40$  & $0.78\pm0.67$ \\
$331.5$ & $0.87\pm0.40$  & $0.55\pm0.99$ \\
$398.5$ & $-0.18\pm0.43$ & $0.90\pm1.51$ \\
$465.5$ & $0.65\pm0.68$  & $0.28\pm2.23$ \\
$532.5$ & $1.03\pm0.72$  & $4.80\pm2.85$ \\
$599.5$ & $0.25\pm0.71$  & $-0.47\pm3.66$ \\
$666.5$ & $1.03\pm0.98$  & $4.97\pm6.56$ \\
\hline
\end{tabular}
\ec 
\end{table}
%<><><><><><><><><><><><><><><><><><><><><><><><><><><><><><><><><><><><><><><><><><><><><><><><><><>%

%<><><><><><><><><><><><><><><><><><><><><><><><><><><><><><><><><><><><><><><><><><><><><><><><><><>%
\begin{figure} 
\bc
\includegraphics[width=8.5cm,clip]{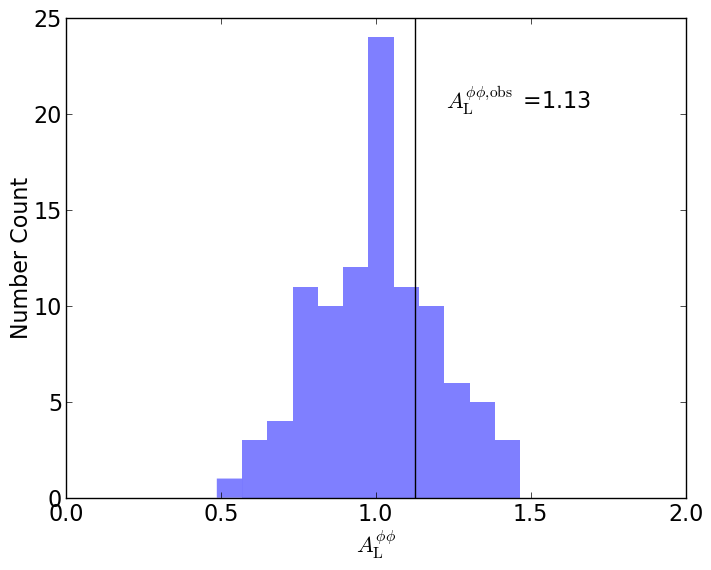}
\caption{
The amplitudes of the cross-spectra of \bk\ and \planck\ 2015 lensing potential maps 
reconstructed from lensed-$\Lambda$CDM+dust+noise simulations (histogram),
and the observed value (vertical line).
}
\label{Fig:A_BKxP}
\ec
\end{figure}
%<><><><><><><><><><><><><><><><><><><><><><><><><><><><><><><><><><><><><><><><><><><><><><><><><><>%

%<><><><><><><><><><><><><><><><><><><><><><><><><><><><><><><><><><><><><><><><><><><><><><><><><><>%
\begin{figure} 
\bc
\includegraphics[width=8.5cm,clip]{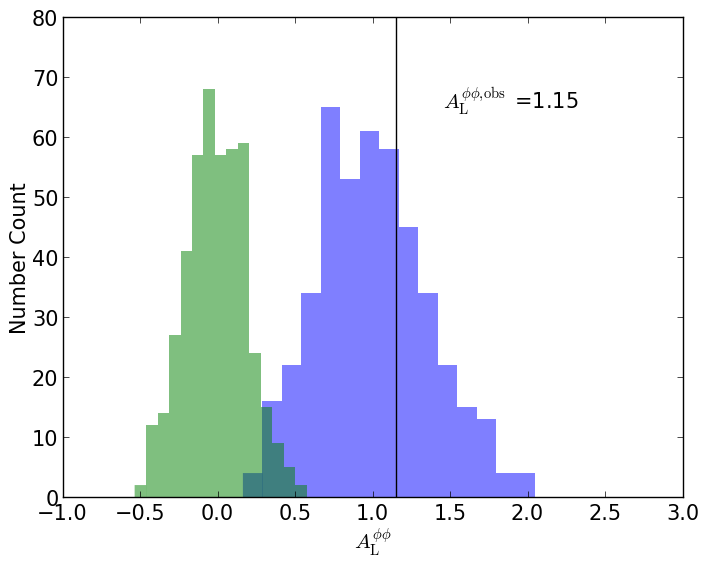}
\caption{
The amplitudes of the auto-spectra of \bk\ lensing potential maps reconstructed from
lensed-$\Lambda$CDM+dust+noise simulations (blue histogram),
and from unlensed-$\Lambda$CDM+dust+noise simulations (green histogram).
The observed value is indicated by the vertical line.
}
\label{Fig:A_BK}
\ec
\end{figure}
%<><><><><><><><><><><><><><><><><><><><><><><><><><><><><><><><><><><><><><><><><><><><><><><><><><>%

\fig{Fig:ClKK} shows the cross-spectrum of the \bk\ and \planck\ lensing-mass fields, 
and the auto-spectrum of the \bk\ data alone. 
Table \ref{table:clkk} shows the bandpowers and $1\sigma$ statistical errors of the lensing power spectrum. 
\fig{Fig:A_BKxP} compares the amplitude of the lensing cross-spectrum 
between \bk\ and \planck\ to lensed-$\Lambda$CDM+dust+noise simulations,
while the line and blue histogram in \fig{Fig:A_BK} do the same thing for the \bk\ auto-spectrum. 
The observed amplitude estimated from the cross-spectrum is $\AL=\mALc\pm\vALc$
and the amplitude estimated from the auto-spectrum is $\AL=\mALa\pm \vALa$.
In each case the uncertainty is taken from the standard deviation of the
lensed-$\Lambda$CDM+dust+noise simulations.
We find that these values are mutually consistent, and are also consistent with the 
\planck\ $\Lambda$CDM expectation within the $1\sigma$ statistical uncertainty. 

To evaluate the rejection significance of the no-lensing hypothesis 
in \fig{Fig:A_BK} we also show the results of a special set of unlensed-$\Lambda$CDM+dust+noise
simulations where there is no sample variance on the lensing component.
Assuming Gaussian statistics we find that the no-lensing hypothesis is rejected at $\sULa\sigma$
which is the highest significance achieved to date using $EB$ lensing estimator.

%Note that the observed value of $\AL$ estimated using the no-lensing simulation is higher than that from 
%the lensed simulation.
%This is because, as shown in Eq.~\eqref{Eq:disconnect}, the disconnected bias includes a contribution from 
%the $B$-mode power spectrum, and in the no-lensing simulation, the $B$-mode power 
%spectrum comes from the instrumental noise alone. 
%Therefore the disconnected bias becomes small in the no-lensing simulation compared to that in the case with lensing.
%This fact makes the significance of rejecting the no-lensing scenario higher compared to that expected  
%only from the variance of the $\AL$ values from simulations. 

The $B$-mode power spectrum can also be used to constrain the amplitude of
the lensing effect and in BKP we quoted the value $\ALB=1.13\pm0.18$ when marginalizing
over $r$ and the dust foreground amplitude.
Updating to the \bk\ spectrum and setting $r=0$ we find $\ALB=\mALb\pm\vALb$.
The agreement of this result with that from the lensing reconstruction
described above verifies that the $B$-mode observed by the \BK\ experiments at
intermediate angular scales is dominated by gravitational lensing.

%<><><><><><><><><><><><><><><><><><><><><><><><><><><><><><><><><><><><><><><><><><><><><><><><><><>%
\begin{figure} 
\bc
\includegraphics[width=8.5cm,clip]{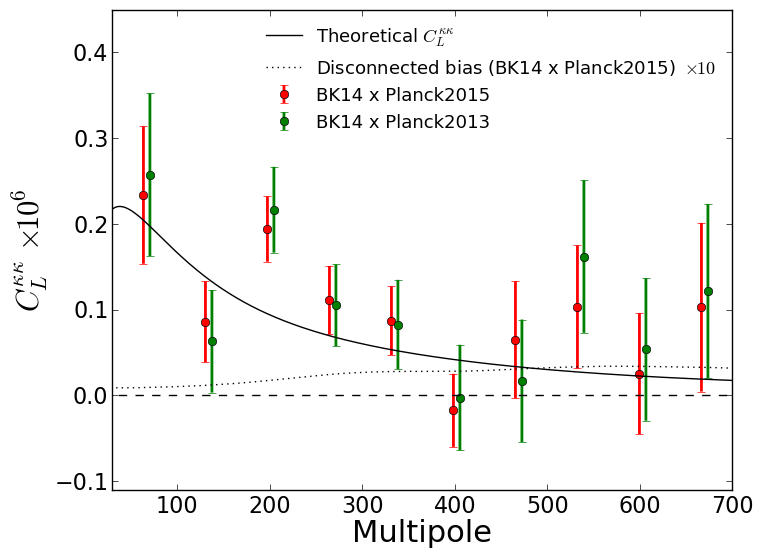}
\caption{
The lensing-mass power spectrum, $C_L^{\kappa\kappa}=L^4C^{\grad\grad}_L/4$, 
estimated from the cross-spectrum between \bk\ and \planck\ 2015 data compared with that 
between \bk\ and \planck\ 2013 data. 
We also show the theoretical expectation of the disconnected bias in the cross-spectrum 
between the \bk\ and \planck\ 2015 data multiplied by $10$. 
The cross-spectrum between \bk\ and \planck\ 2013 is offset in $L$ for clarity. 
}
\label{Fig:planck}
\ec
\end{figure}
%<><><><><><><><><><><><><><><><><><><><><><><><><><><><><><><><><><><><><><><><><><><><><><><><><><>%

To show that the disconnected bias in the cross-spectrum is small,
an analytic estimate multiplied by 10$\times$ for clarity is compared
in \fig{Fig:planck} to the \bk/\planck\ 2015 cross-spectrum.
The inclusion of this bias changes the value of the lensing amplitude by less than $1\%$.
In addition, we show an alternate cross-spectrum taken between \bk\ and the \planck\ 2013 data. 
As mentioned earlier, the \bk\ and \planck\ 2013 cross-spectrum is free of any disconnected bias.
Therefore, the similarity of these two spectra also suggests that the disconnected bias in
the \bk\ and \planck\ 2015 cross-spectrum is small.

%////////////////////////////////////////////////////////////////////////////////////////////////////%
\section{Consistency checks and null tests} \label{systematics}
%////////////////////////////////////////////////////////////////////////////////////////////////////%

In this section, we discuss systematics in the reconstructed lensing potential. 
$B$-modes in the $EB$ estimator for $\grad$ are an order of magnitude fainter than the $E$-modes, 
and need to be tested for non-negligible contributions from systematics or leakage from $E$-modes. 
The matrix-purified \bk\ $E$- and $B$-modes up to $\l\simeq350$ used in this paper have already
passed the long list of systematics and null tests described in BK-I, BK-III and BK-VI.
In the baseline results presented above we include additional modes
up to $\lmax=700$, and we see below that the modes in the range $350<\ell<600$
carry a significant portion of the total available statistical weight.
In this section we therefore discuss additional tests that demonstrate the robustness of the 
reconstructed $\grad$ map and the lensing spectrum. 
Furthermore, note that the cross-spectrum of \bk\ and \planck, which produces the most stringent constraint
on $\AL$ in this paper, is immune to additive bias from all known systematics, 
and is highly insensitive to the dust foreground. 

%::::::::::::::::::::::::::::::::::::::::::::::::::::::::::::::::::::::::::::::::::::::::::::::::::::%
\subsection{Null tests} 
%::::::::::::::::::::::::::::::::::::::::::::::::::::::::::::::::::::::::::::::::::::::::::::::::::::%

%<><><><><><><><><><><><><><><><><><><><><><><><><><><><><><><><><><><><><><><><><><><><><><><><><><>%
\begin{figure} 
\bc
\includegraphics[width=8.5cm,clip]{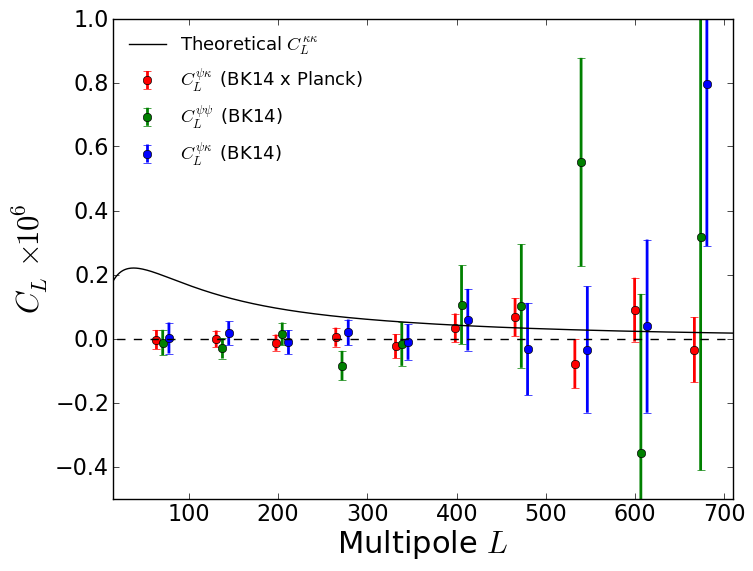}
\caption{
{\it Curl null test}: 
The cross-spectrum of the curl-mode reconstructed from \bk\ data 
and the \planck\ lensing-mass field (red), the auto-spectrum of the \bk\ curl-mode (green), 
and the cross-spectrum of the \bk\ lensing-mass field and curl-mode (blue). 
For comparison, we also show the theoretical lensing-mass power spectrum (black).
The power spectra are offset in $L$ for clarity.
}
\label{Fig:curl}
\ec
\end{figure}
%<><><><><><><><><><><><><><><><><><><><><><><><><><><><><><><><><><><><><><><><><><><><><><><><><><>%

%<><><><><><><><><><><><><><><><><><><><><><><><><><><><><><><><><><><><><><><><><><><><><><><><><><>%
\begin{figure} 
\bc
\includegraphics[width=7.5cm,clip]{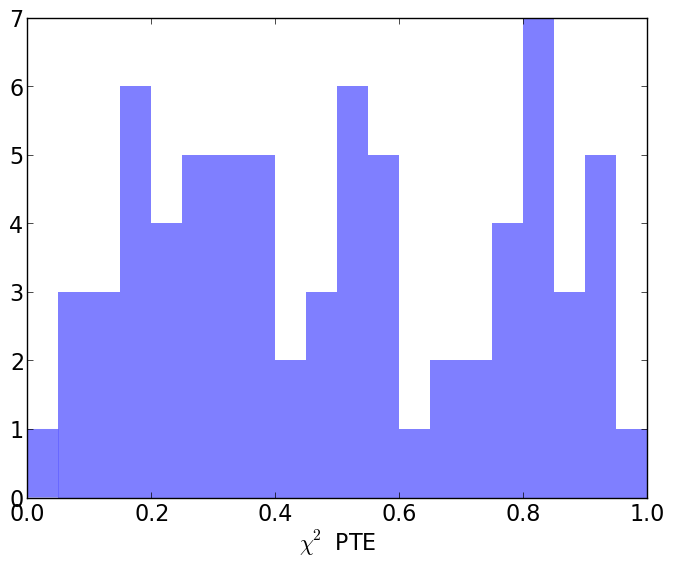}
\caption{
Distribution of the jackknife $\chi^2$ PTE values. 
}
\label{Fig:null}
\ec
\end{figure}
%<><><><><><><><><><><><><><><><><><><><><><><><><><><><><><><><><><><><><><><><><><><><><><><><><><>%

%<><><><><><><><><><><><><><><><><><><><><><><><><><><><><><><><><><><><><><><><><><><><><><><><><><>%
\begin{table}
\renewcommand{\arraystretch}{1.8}
\bc
\caption{
Probability to exceed a $\chi^2$ statistic for the curl null test and the jackknife tests. 
}
\label{table:PTE}
\begin{tabular}{l|cc|ccc} \hline
& \multicolumn{2}{c|}{\bk\ $\times$ \planck} & \multicolumn{3}{c}{\bk} \\ \hline
& $\grad\times\grad$ & $\curl\times\grad$ & $\grad\times\grad$ & $\curl\times\grad$ & $\curl\times\curl$ \\ \hline
Curl                    & ---  & 0.77 & ---  & 0.92 & 0.34 \\ 
Deck                    & 0.51 & 0.48 & 0.34 & 0.06 & 0.12 \\
Scan Dir                & 0.37 & 0.43 & 0.87 & 0.15 & 0.57 \\
Tag Split               & 0.30 & 0.85 & 0.36 & 0.86 & 0.73 \\
Tile                    & 0.30 & 0.16 & 0.20 & 0.05 & 0.54 \\
Phase                   & 0.69 & 0.68 & 0.92 & 0.76 & 0.25 \\
Mux Col                 & 0.18 & 0.22 & 0.46 & 0.38 & 0.35 \\
Alt Deck                & 0.18 & 0.72 & 0.39 & 0.16 & 0.16 \\
Mux Row                 & 0.49 & 0.80 & 0.60 & 0.58 & 0.09 \\
Tile/Deck               & 0.20 & 0.36 & 0.83 & 0.84 & 0.88 \\
Focal Plane inner/outer & 0.09 & 0.12 & 0.41 & 0.35 & 0.28 \\
Tile top/bottom         & 0.84 & 0.51 & 0.51 & 0.77 & 0.28 \\
Tile inner/outer        & 0.31 & 0.05 & 0.91 & 0.64 & 0.78 \\
Moon                    & 0.02 & 0.84 & 0.18 & 0.53 & 0.83 \\
A/B offset best/worst   & 0.93 & 1.00 & 0.57 & 0.24 & 0.59 \\
\hline
\end{tabular}
\ec 
\end{table}
%<><><><><><><><><><><><><><><><><><><><><><><><><><><><><><><><><><><><><><><><><><><><><><><><><><>%

In the following we present results of i) a curl-null test, and ii) jackknife tests, 
which are expected to be consistent with zero unless there are systematics remaining in the data. 

To test this quantitatively, we use the probability to exceed (PTE) the value of $\chi^2$ 
obtained from observations, under the assumption that the fiducial power spectrum is zero 
in all multipole bins. 
The PTE is evaluated from the simulation set with the same method as in the BK-I paper.
Table~\ref{table:PTE} summarizes the PTE values obtained. 
\fig{Fig:null} shows the distribution of the jackknife $\chi^2$ PTE.

\subsubsection{Curl null test} 

The curl-mode is mathematically similar to lensing but cannot be generated by scalar perturbations at linear order. 
As described in \sec{analysis}, the curl-mode is estimated by replacing the weight function with $w_{\bl,\bL}^\curl$, 
and the reconstruction noise level in the curl-mode is similar to that in the lensing potential. 
It is therefore commonly used as an important check for any residual systematics in
lensing reconstruction analysis. 

\fig{Fig:curl} shows the cross-spectrum between the \bk\ curl-mode and the \planck\ lensing potential, 
the \bk\ curl-mode auto-spectrum, and the cross-spectrum between the \bk\ lensing potential and curl-mode.
For illustrative purposes,
similar to the relationship between $\kappa$ and $\grad$, we define $\psi_{\bL}=-L^2\curl_{\bL}/2$, 
and show the power spectrum of $\psi$ instead of $\curl$. 
We compute the corresponding PTEs for these power spectra (see Table~\ref{table:PTE}), 
finding no evidence of systematics in these curl-null tests.

\subsubsection{Jackknife tests}

As part of our standard data reduction we form multiple pairs of jackknife
maps which split the data into approximately equal halves, and which
should contain (nearly) identical sky signal, but which might be expected
to contain different systematic contamination.
We then difference these pairs of maps and search for signals which
are inconsistent with the noise expectation---see BK-I, BK-III and BK-VI
for further details.
Here we take these jackknife maps, perform the lensing reconstruction on them, 
and as usual look for signals which are inconsistent with null.

Table~\ref{table:PTE} gives the PTE values.
We find no evidence of spurious signals in the lensing potential.

%::::::::::::::::::::::::::::::::::::::::::::::::::::::::::::::::::::::::::::::::::::::::::::::::::::%
\subsection{Consistency checks} 
%::::::::::::::::::::::::::::::::::::::::::::::::::::::::::::::::::::::::::::::::::::::::::::::::::::%

%<><><><><><><><><><><><><><><><><><><><><><><><><><><><><><><><><><><><><><><><><><><><><><><><><><>%
\begin{table}
\renewcommand{\arraystretch}{1.8}
\bc
\caption{
The amplitude of the lensing potential power spectrum $\AL$ estimated with different analysis choices.
}
\label{table:AL}
\begin{tabular}{lcc} \hline 
 & \bk\ $\times$ \planck & \bk \\ \hline 
Baseline & $\mALc\pm\vALc$ & $\mALa\pm\vALa$ \\ \hline 
$\lmax=650$            & $1.07\pm0.20$ & $1.21\pm0.36$ \\
$\lmax=600$            & $1.14\pm0.20$ & $1.26\pm0.36$ \\
$\lmax=350$            & $1.41\pm0.30$ & $1.97\pm0.84$ \\
$\lmin=150$            & $1.13\pm0.20$ & $1.14\pm0.36$ \\
$\lmin=200$            & $1.07\pm0.20$ & $0.95\pm0.40$ \\
$\lmax^{\rm B}=350$    & $1.24\pm0.22$ & $1.33\pm0.45$ \\
Diff. beam ellipticity & $1.11\pm0.20$ & $1.14\pm0.36$ \\
Apodization            & $1.07\pm0.22$ & $0.99\pm0.39$ \\
\hline
\end{tabular}
\ec 
\end{table}
%<><><><><><><><><><><><><><><><><><><><><><><><><><><><><><><><><><><><><><><><><><><><><><><><><><>%

As consistency checks of the \bk\ lensing potential, we calculate the lensing power spectrum 
while varying the following analysis choices from their baseline values, and give the resulting
alternate values of $\AL$ in Table~\ref{table:AL}.

\bi
\item Maximum multipole: $\lmax$ \\
In our baseline analysis, the nominal maximum multipole of the $E$- and $B$-modes used for 
the lensing reconstruction in \eq{Eq:def-AXY} is $\lmax=700$.
Reducing the value of $\lmax$ to $650$ and $600$ we see small changes in the constraint on $\AL$. 
However, if we reduce $\lmax$ to $350$ to match the range probed by jackknife tests in BK-VI, 
the values of $\AL$ shift up, and the statistical errors increase.
To quantify how likely the up-shifts are to occur by chance 
we compute the corresponding shifts when making the same change in the
simulations, and find a positive shift greater than the observed one
$10$\% of the time for the cross-spectrum and $15$\% of the time for the auto-spectrum.
\item Minimum multipole: $\lmin$ \\
For the baseline analysis the minimum multipole of the $E$-modes is set to $30$ in \eq{Eq:def-AXY}
(due to the timestream filtering), while the minimum multipole of the $B$-modes is set to $150$
(to ensure that the contributions of the dust foreground is small compared to the noise and lensing signal).
Raising $\lmin$ for the $E$-modes to $150$ we see very small changes to the $\AL$ results,
while raising both to $200$ we see modest changes.
\item Maximum multipole of the $B$-mode polarization: $\lmax^{\rm B}$ \\
As mentioned above, $B$-modes are used up to a nominal $\lmax=700$ in our baseline analysis. 
The $B$-mode polarization at $\l\gtrsim 350$ is not as well tested against various systematics.
However, unlike $E$-modes, $B$-modes at smaller scales $\l>350$ do not contribute significantly in estimating $\AL$. 
We repeat the analysis removing $B$-modes at $\l>350$, 
and find only a moderate change in the results and their statistical uncertainties.
\item Differential beam ellipticity: \\
In our pair-differencing analysis differences in the beam shapes between the A and B detectors
of each pair generates temperature-to-polarization leakage. 
We filter out the leading order modes of this leakage using a technique
which we call deprojection (see BK-III for details). 
For differential beam ellipticity, however, we do not use deprojection because it introduces a bias in 
$TE$. Instead, in our standard analysis, we subtract the expected temperature-to-polarization leakage based on the
measured differential beam ellipticity.
To test whether the lensing results are sensitive to differential beam ellipticity, we repeat 
the lensing reconstruction from maps {\em without} this subtraction and find only a very small change in the results.
\item Apodization: \\
To mitigate the noisy regions around the survey boundary, after obtaining the purified $E$ and $B$ modes, 
our standard analysis applies an inverse variance apodization window. 
We also perform the analysis using the sine apodization defined in \citet{Namikawa:2013}
and find only a small change in the results.
\ei

%::::::::::::::::::::::::::::::::::::::::::::::::::::::::::::::::::::::::::::::::::::::::::::::::::::%
\subsection{Effects of beam systematics} 
%::::::::::::::::::::::::::::::::::::::::::::::::::::::::::::::::::::::::::::::::::::::::::::::::::::%

%<><><><><><><><><><><><><><><><><><><><><><><><><><><><><><><><><><><><><><><><><><><><><><><><><><>%
\begin{figure} 
\bc
\includegraphics[width=8.5cm,clip]{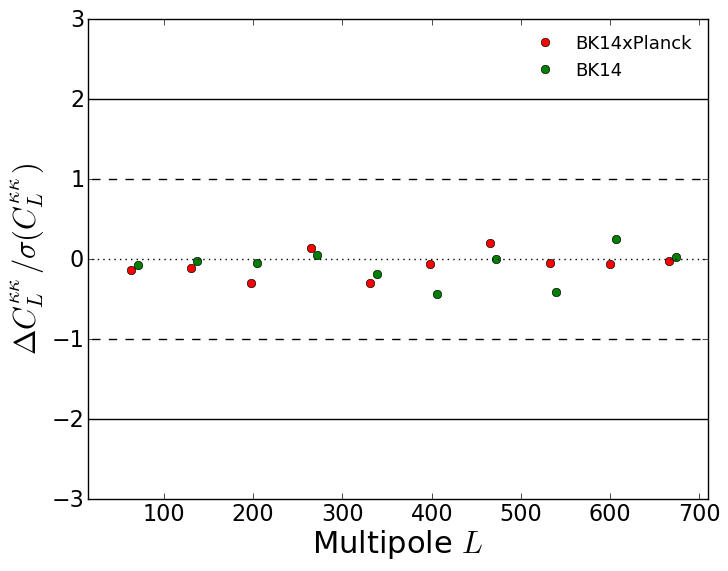}
\caption{
The difference of the lensing power spectrum when subtracting from the \bk\ $Q/U$ maps 
a nominal undeprojected residual as derived from per-channel beam maps
(red: BK14$\times$\planck, blue:BK14), divided by the $1\,\sigma$ statistical uncertainty. 
}
\label{Fig:beam}
\ec
\end{figure}
%<><><><><><><><><><><><><><><><><><><><><><><><><><><><><><><><><><><><><><><><><><><><><><><><><><>%

Beam shape mismatch of each detector pair leads to a leakage from the bright
temperature anisotropies into polarization (e.g. \citealt{Hu:2003,Miller:2008,Su:2009}). 
In our analysis, this leakage is
mitigated by deprojecting (or for ellipticity, subtracting) several modes
corresponding approximately
to the difference of two elliptical Gaussians (see BK-III for
details). To assess the level of leakage remaining after deprojection, we use
calibration data consisting of high precision, per-detector beam maps described
in BK-IV. In special simulations, we explicitly convolve these beam maps onto an
input $T$ sky and process the resulting simulated timestream in the normal
manner, including deprojection, to produce maps of the ``undeprojected residual.'' 
In BK-V, this residual was treated as an upper limit to possible
residual systematics. Here, as an additional check, we try subtracting this
nominal residual from the maps and re-extracting the lensing potential. 
\fig{Fig:beam} shows the differences in the resulting spectrum in units of the bandpower
uncertainties, finding that the difference is small compared to the $1\,\sigma$
statistical uncertainty.

In addition to temperature-to-polarization leakage caused by beam mismatch, beam
asymmetry as well as detector-to-detector beam shape variation can produce a
spurious lensing signal if non-uniform map coverage leads to an effective beam
that is spatially dependent (e.g. \citealt{P13:phi}). 
The beam map simulation procedure described above
does not probe this effect in the $EB$ estimator because the input maps do not
contain polarization. However, we note that ellipticity is the dominant
component of beam asymmetry and beam shape variation in \bicep2 and \keck~(see
BK-IV, Table~2). We also note that beam ellipticity is a strong function of
radial position in the focal plane (see BK-IV, Figures~12-13), so that the focal
plane inner/outer jackknife listed in Table~\ref{table:PTE} is a good proxy for
a beam ellipticity jackknife. The fact that this null test passes limits the
contribution from beam asymmetry and beam variation to less than the uncertainty. 

We finally test the effects of the beam correction to $E$-/$B$-modes based on the observed level of 
temperature anisotropies at high $\l$ (\sec{analysis}). 
We repeat the same lensing reconstruction without the beam correction, 
and estimate the lensing amplitude. 
We find that $\AL$ increases while the statistical error is unchanged compared to the baseline results, 
and the differences of $\AL$ are $\Delta\AL=0.05$ for \bk $\times$ \planck\ and $\Delta\AL=0.15$ for \bk. 
These changes are within the $1\sigma$ statistical error.

%::::::::::::::::::::::::::::::::::::::::::::::::::::::::::::::::::::::::::::::::::::::::::::::::::::%
\subsection{Effects of absolute calibration error}
%::::::::::::::::::::::::::::::::::::::::::::::::::::::::::::::::::::::::::::::::::::::::::::::::::::%

Although the lensing potential is in principle a dimensionless quantity, the measured lensing potential 
depends on the overall amplitude of the polarization map. 
The calibration uncertainties in $E$- and $B$-modes therefore propagate into an error 
in the amplitude of the lensing potential spectrum (e.g. \citealt{PB14:phi}). 
The absolute calibration uncertainty, $\delta$, is $1.3$\% in the \bicep2/\keck\
polarization maps (BK-I). 
Given this uncertainty on amplitudes of the $E$- and $B$-modes, 
the resultant systematic uncertainties in the lensing spectral amplitudes are $\Delta\AL=4\delta=0.052$
for the \bk\ auto-spectrum and $\Delta\AL=2\delta=0.026$ for the cross-spectrum with \planck, significantly smaller 
than the statistical uncertainties. 
Since the estimate of the curl-mode power spectrum is also affected in the same manner,
non-detection of the curl mode also indicates that the effect of these uncertainties is 
negligible compared to the statistical errors.

%////////////////////////////////////////////////////////////////////////////////////////////////////%
\section{Conclusions} \label{conclusions}
%////////////////////////////////////////////////////////////////////////////////////////////////////%

In this paper we have reconstructed the lensing potential from the \bk\ polarization data,
and taken its cross-spectrum with the public \planck\ lensing potential,
as well as the auto-spectrum of the \bk\ alone.
The amplitude of the cross-spectrum with \planck\ is constrained to be
$\AL=\mALc\pm\vALc$, while the auto-spectrum has amplitude $\AL=\mALa\pm\vALa$.
By comparing the auto-spectrum to special unlensed simulations we reject the no-lensing hypothesis at
$\sULa\sigma$ significance, which is the highest significance achieved to date using $EB$ lensing estimator.
We have performed several consistency checks and null tests, and 
find no evidence for spurious signals in our reconstructed map and spectra.

This paper demonstrates for the first time lensing reconstruction using $B$-modes in the intermediate multipole 
range. The results verify that the $B$-mode power observed by the \BK\ experiments on 
these intermediate angular scales is dominated by gravitational lensing. 
The good agreement between these results and $\ALB=\mALb\pm\vALb$ from the \bk\ $B$-mode spectrum starts 
to place constraints on any alternative sources of $B$-modes at these angular scales,
such as cosmic strings (e.g., \citealt{Seljak:2006,Pogosian:2008}), 
primordial magnetic fields (e.g., \citealt{Shaw:2010,Bonvin:2014})
and cosmic birefringence induced by interaction between a massless pseudo-scalar field and photons
(e.g., \citealt{Pospelov:2009,Lee:2015,PB15:rot}). The calculation of formal quantitative constraints is 
rather involved and depends on the assumed statistical properties of the alternative $B$-mode sources. 
We leave that to future work. 

Looking ahead, the reconstructed lensing potential can be used to cross-correlate with other astronomical tracers. 
However, the reconstruction noise of the \BK\ data will limit its usefulness as a cosmological probe in the 
era of DES \citep{DES}, DESI \citep{DESI}, and LSST \citep{LSST}. 
As the sensitivity of \BK\ improves, our main objective is to use a well-measured 
deflection map $\grad$ to form a degree-scale $B$-mode lensing template, which can then be used to 
improve our final uncertainties on $r$ (i.e., ``delensing''). 
Multiple studies have shown that high resolution CMB polarization data
(e.g., \citealt{Seljak:2003pn,Smith:2010gu}), the CIB \citep{Simard:2015,Sherwin:2015}, 
galaxy clustering \citep{Namikawa:2016}, or weak lensing \citep{Sigurdson:2005cp,Marian:2007} 
can all improve measurements of $\grad$.
In addition to the lensing potential presented here there already exists in the \BK\
field data from the \planck\ CIB measurements \citep{P13:CIBxphi,P13:CIB},
as well as high resolution CMB maps from \sptpol.
We are exploring the formation of a lensing template using an optimal combination
of these and anticipate using this in our likelihood analysis in the near future.
This template will considerably improve as \sptIIIg \citep{SPT3G} comes online.

%////////////////////////////////////////////////////////////////////////////////////////////////////%
% BACK MATTER 
%////////////////////////////////////////////////////////////////////////////////////////////////////%

% Acknowledgments %
\acknowledgements
% To keep this under control:
% Only financial support from funding agencies--not institutional.
% No personal grants, fellowships, or support acknowledgements.

The \keckarray\ project has been made possible through support from the National Science Foundation 
under Grants ANT-1145172 (Harvard), ANT-1145143 (Minnesota) \& ANT-1145248 (Stanford), and from the 
Keck Foundation (Caltech). The development of antenna-coupled detector technology was supported
by the JPL Research and Technology Development Fund and Grants No.\ 06-ARPA206-0040 and 10-SAT10-0017 
from the NASA APRA and SAT programs. The development and testing of focal planes were supported
by the Gordon and Betty Moore Foundation at Caltech. Readout electronics were supported by a Canada 
Foundation for Innovation grant to UBC. The computations in this paper were run on the Odyssey cluster
supported by the FAS Science Division Research Computing Group at Harvard University. The analysis 
effort at Stanford and SLAC is partially supported by the U.S. Department of Energy Office of Science.
We thank the staff of the U.S. Antarctic Program and in particular the South Pole Station without 
whose help this research would not have been possible. Most special thanks go to our heroic 
winter-overs Robert Schwarz and Steffen Richter. We thank all those who have contributed past efforts 
to the \bicep--\keckarray\ series of experiments, including the \bicepone\ team.
T.~N. acknowledges support from Japan Society for the Promotion of Science Postdoctoral Fellowships for Research Abroad.

% Appendix %
\appendix
%////////////////////////////////////////////////////////////////////////////////////////////////////%
\section{Disconnected bias estimation} \label{sec:disconnected}
%////////////////////////////////////////////////////////////////////////////////////////////////////%

The realization-dependent method for the disconnected bias given in Eq.~\eqref{Eq:hN} comes 
naturally from deriving the optimal estimator for the lensing-induced trispectrum.
Here we briefly summarize derivation of Eq.~\eqref{Eq:hN} 
(see appendix A of \citealt{Namikawa:2013} for a thorough derivation). 

The lensing-induced trispectrum which is relevant to our analysis is given by 
(see e.g. \citealt{Lewis:2006fu})
%----------------------------------------------------------------------------------------------------%
\al{
	T_{\bl_1\bl_2\bl_3\bl_4} 
	\equiv \ave{E_{\bl_1}B_{\bl_2}E_{\bl_3}B_{\bl_4}}_{\rm C} 
%		\notag \\
	\simeq \delta^D_{\bl_1+\bl_2+\bl_3+\bl_4}
		[w^{\grad}_{\bl_1+\bl_2,\bl_1} w^{\grad}_{\bl_3+\bl_4,\bl_3} C_{|\bl_1+\bl_2|}^{\grad\grad}
		+ w^{\grad}_{\bl_1+\bl_4,\bl_1} w^{\grad}_{\bl_2+\bl_3,\bl_2} C_{|\bl_1+\bl_4|}^{\grad\grad}]
	\,,
}
%----------------------------------------------------------------------------------------------------%
where $w^{\grad}_{\bL,\bl}$ is given in Eq.~\eqref{Eq:weight:g}, $C_L^{\grad\grad}$ is the lensing potential 
power spectrum, and $\delta^D_\bL$ is the Dirac delta function in Fourier space. 
In the Edgeworth expansion of the $E$- and $B$-mode likelihood, the term containing the above trispectrum 
is given by \citep{Regan:2010}
%----------------------------------------------------------------------------------------------------%
\al{
	\mC{L} \propto \left[\prod_{i=1}^4\Int{2}{\bl_i}{(2\pi)^2}\right] T_{\bl_1\bl_2\bl_3\bl_4}
		\PD{}{E_{\bl_1}}\PD{}{B_{\bl_2}}\PD{}{E_{\bl_3}}\PD{}{B_{\bl_4}} \mC{L}_{\rm g}
	\,, 
}
%----------------------------------------------------------------------------------------------------%
where $\mC{L}_{\rm g}$ is the Gaussian likelihood of the $E$- and $B$-mode:
%----------------------------------------------------------------------------------------------------%
\al{
	\mC{L}_{\rm g} \propto \exp \left(-\frac{1}{2}\left[\prod_{i=1}^2\Int{2}{\bl_i}{(2\pi)^2}\right]\sum_{a,b=E,B}
		a_{\bl_1}\{\bR{C}^{-1}\}^{a_{\bl_1}b_{\bl_2}}b_{\bl_2}\right) \,.
}
%----------------------------------------------------------------------------------------------------%
Here $\{\bR{C}\}^{a_{\bl}b_{\bl}}=\ave{a_{\bl}b_{\bl'}}$ is the covariance matrix, and we omit the 
normalization of the above Gaussian likelihood. 

The optimal estimator for the lensing power spectrum in the trispectrum is obtained by maximizing 
the CMB likelihood. The approximate formula which is numerically tractable is proportional to 
the derivative of the log-likelihood with respect to $C_L^{\grad\grad}$. 
The derivative of the above likelihood with respect to the lensing potential power spectrum is given by \citep{Namikawa:2013}
%----------------------------------------------------------------------------------------------------%
\al{
	\PD{\mC{L}}{C^{\grad\grad}_L} 
		\propto \left[\prod_{i=1}^2\Int{2}{\bl_i}{(2\pi)^2}\right]
		w^\grad_{\bL,\bl_1} w^\grad_{-\bL,\bl_2}
		\PD{}{E_{\bl_1}}\PD{}{B_{\bL-\bl_1}} \PD{}{E_{\bl_2}}\PD{}{B_{-\bL-\bl_2}} \mC{L}_{\rm g}
	%\notag \\
	%	&\simeq \Int{2}{\bl_1}{(2\pi)^2}w^{\grad}_{\bL,\bl_1}
	%	\PD{}{E_{\bl_1}}\PD{}{B_{\bL-\bl_1}} (\uestg^{EB}_{-\bL}-\ave{\uestg^{EB}_{-\bL}}) \mC{L}_{\rm g}
	%\notag \\
		\simeq \left(|\uestg^{EB}_{\bL}|^2 - \frac{\hN^\grad_{\bL}}{(A^\grad_{\bL})^2} \right) \mC{L}_{\rm g}
	\,. 
}
%----------------------------------------------------------------------------------------------------%
After correcting the normalization for the unbiased estimator, the above equations leads to 
Eq.~\eqref{Eq:est-clgg}. 

Realization-dependent methods are useful to suppress spurious off-diagonal elements 
in the covariance matrix of the power spectrum estimates (e.g., \citealt{Dvorkin:2009,Hanson:2010rp}). 
As discussed in \citet{Namikawa:2012pe}, 
the disconnected bias estimation described above is less sensitive to errors in covariance 
compared to the other approaches. 
To see this, using Eq.~\eqref{Eq:est}, we rewrite Eq.~\eqref{Eq:disconnect} as
%----------------------------------------------------------------------------------------------------%
\al{
	\hN_{\bL}^{\grad} 
	&= (A^\grad_{\bL})^2\Int{2}{\bl}{(2\pi)^2} \Int{2}{\bl'}{(2\pi)^2} w^{\grad}_{\bL,\bl}w^{\grad}_{-\bL,\bl'}
		\bigg[\ol{\bR{C}}^{\rm EE}_{\bl,\bl'}\bB_{\bL-\bl}\bB_{-\bL-\bl'}
		+\ol{\bR{C}}^{\rm BB}_{\bL-\bl,-\bL-\bl'}\bE_{\bl}\bE_{\bl'}
		-\ol{\bR{C}}^{\rm EE}_{\bl,\bl'}\ol{\bR{C}}^{\rm BB}_{\bL-\bl,-\bL-\bl'}
	\notag \\
	&\qquad 
		+\ol{\bR{C}}^{\rm EB}_{\bl',\bL-\bl}\bE_{\bl}\bB_{-\bL-\bl'}
		+\ol{\bR{C}}^{\rm EB}_{\bl,-\bL-\bl'}\bE_{\bl'}\bB_{\bL-\bl}
		-\ol{\bR{C}}^{\rm EB}_{\bl,-\bL-\bl'}\ol{\bR{C}}^{\rm EB}_{\bl',\bL-\bl}\bigg]
	\,. \label{Eq:hN}
}
%----------------------------------------------------------------------------------------------------%
For example, replacing the covariance matrix with an incorrect covariance model, 
$\ol{\bR{C}}^{\rm EE}+\Sigma^{\rm EE}$, we obtain
%----------------------------------------------------------------------------------------------------%
\al{
	\hN_{\bL}^{\grad}
	&= (A^\grad_{\bL})^2\Int{2}{\bl}{(2\pi)^2} \Int{2}{\bl'}{(2\pi)^2} w^{\grad}_{\bL,\bl}w^{\grad}_{-\bL,\bl'}
		\Sigma^{\rm EE}_{\bl,\bl'}(\bB_{\bL-\bl}\bB_{-\bL-\bl'}-\ol{\bR{C}}^{\rm BB}_{\bL-\bl,-\bL-\bl'})
		+ \mC{O}([\Sigma^{\rm EE}]^2)
	\,. \label{Eq:hN}
}
%----------------------------------------------------------------------------------------------------%
and $\ave{\hN_{\bL}^{\grad}}$ has no contribution from $\mC{O}(\Sigma^{\rm EE})$. 

Note that the estimators for $C_L^{\curl\curl}$ and $C_L^{\grad\curl}$ are also derived in the same way \citep{Namikawa:2013}. 
The estimator for the curl-mode power spectrum is given by replacing $\estg_{\bL}$ with $\estc_{\bL}$, 
while the disconnected bias for $C_L^{\grad\curl}$ is estimated from
%----------------------------------------------------------------------------------------------------%
\al{
	\hN^{\grad\curl}_{\bL} = 
		\ave{\Re[(\estg^{E_{\bm 1},\hB}_{\bL}+\estg^{\hE,B_{\bm 1}}_{\bL})
		(\estc^{E_{\bm 1},\hB}_{\bL}+\estc^{\hE,B_{\bm 1}}_{\bL})^*]}_{\bm 1}
		- \frac{1}{2}\ave{\Re[(\estg^{E_{\bm 1},B_{\bm 2}}_{\bL}+\estg^{E_{\bm 2},B_{\bm 1}}_{\bL})
		(\estc^{E_{\bm 1},B_{\bm 2}}_{\bL}+\estc^{E_{\bm 2},B_{\bm 1}}_{\bL})^*]}_{\bm 1,2}
	\,.
}
%----------------------------------------------------------------------------------------------------%

% References %
\bibliographystyle{apj}
\bibliography{cite}

\begin{thebibliography}{}
\expandafter\ifx\csname natexlab\endcsname\relax\def\natexlab#1{#1}\fi

\bibitem[{Abazajian {et~al.}(2015)}]{Abazajian:2013oma}
Abazajian, K.~N., {et~al.} 2015, Astropart. Phys., 63, 66

\bibitem[{Allison {et~al.}(2015)}]{Allison:2015}
Allison, R., {et~al.} 2015, \prd, 92, 123535

\bibitem[{Benson {et~al.}(2014)}]{SPT3G}
Benson, B.~A., {et~al.} 2014, arXiv:1407.2973

\bibitem[{{\textsc{Bicep2} and {\it Planck} Collaborations}(2015)}]{BKP}
{\textsc{Bicep2} and {\it Planck} Collaborations}. 2015, \prl, 114, 101301

\bibitem[{{\textsc{Bicep2} Collaboration I}(2014)}]{B2I}
{\textsc{Bicep2} Collaboration I}. 2014, \prl, 112, 241101

\bibitem[{{\textsc{Bicep2} Collaboration II}(2014)}]{B2II}
{\textsc{Bicep2} Collaboration II}. 2014, \apj, 792, 62

\bibitem[{{\textsc{Bicep2} Collaboration III}(2015)}]{B2III}
{\textsc{Bicep2} Collaboration III}. 2015, \apj, 814, 110

\bibitem[{{\textsc{Bicep2} Collaboration IV}(2015)}]{BKIV}
{\textsc{Bicep2} Collaboration IV}. 2015, \apj, 806, 206

\bibitem[{{\textsc{Bicep2} / {\it Keck Array} Collaborations V}(2015)}]{BKV}
{\textsc{Bicep2} / {\it Keck Array} Collaborations V}. 2015, \apj, 126, 811

\bibitem[{{\textsc{Bicep2} / {\it Keck Array} Collaboration VI}(2015)}]{BKVI}
{\textsc{Bicep2} / {\it Keck Array} Collaboration VI}. 2015, \prl, 116, 031302

\bibitem[{{\textsc{Bicep2} / {\it Keck Array} Collaboration VII}(2016)}]{BKVII}
{\textsc{Bicep2} / {\it Keck Array} Collaboration VII}. 2016, arXiv:1603.05976

\bibitem[{Bonvin {et~al.}(2014)Bonvin, Durrer, \& Maartens}]{Bonvin:2014}
Bonvin, C., Durrer, R., \& Maartens, R. 2014, \prl, 112, 191303

\bibitem[{Cooray {et~al.}(2005)Cooray, Kamionkowski, \&
  Caldwell}]{Cooray:2005hm}
Cooray, A., Kamionkowski, M., \& Caldwell, R.~R. 2005, \prd, 71, 123527

\bibitem[{Das {et~al.}(2011)}]{Das:2011}
Das, S., {et~al.} 2011, \prl, 107, 021301

\bibitem[{Das {et~al.}(2014)}]{Das:2014}
---. 2014, \jcap, 04, 014

\bibitem[{Dvorkin \& Smith(2009)}]{Dvorkin:2009}
Dvorkin, C., \& Smith, K.~M. 2009, \prd, 79, 043003

\bibitem[{Giannantonio {et~al.}(2015)}]{DESxPlanck}
Giannantonio, T., {et~al.} 2015, \mnras, 456, 3213

\bibitem[{Hanson {et~al.}(2011)Hanson, Challinor, Efstathiou, \&
  Bielewicz}]{Hanson:2010rp}
Hanson, D., Challinor, A., Efstathiou, G., \& Bielewicz, P. 2011, \prd, 83,
  043005

\bibitem[{Hanson {et~al.}(2010)Hanson, Challinor, \& Lewis}]{Hanson:review}
Hanson, D., Challinor, A., \& Lewis, A. 2010, Gen. Relativ. Gravit., 42, 2197

\bibitem[{Hanson \& Lewis(2009)}]{Hanson:2009gu}
Hanson, D., \& Lewis, A. 2009, \prd, 80, 063004

\bibitem[{Hanson {et~al.}(2009)Hanson, Rocha, \& Gorski}]{Hanson:2009}
Hanson, D., Rocha, G., \& Gorski, K. 2009, \mnras, 400, 2169

\bibitem[{Hanson {et~al.}(2013)}]{Hanson:2013daa}
Hanson, D., {et~al.} 2013, \prl, 111, 141301

\bibitem[{Hirata \& Seljak(2003{\natexlab{a}})}]{Hirata:2003}
Hirata, C.~M., \& Seljak, U. 2003{\natexlab{a}}, \prd, 67, 043001

\bibitem[{Hirata \& Seljak(2003{\natexlab{b}})}]{Hirata:2003ka}
---. 2003{\natexlab{b}}, \prd, 68, 083002

\bibitem[{Hu(2001)}]{Hu:2001}
Hu, W. 2001, \apj, 557, L79

\bibitem[{Hu(2002)}]{Hu:2001a}
---. 2002, \prd, 65, 023003

\bibitem[{Hu {et~al.}(2003)Hu, Hedman, \& Zaldarriaga}]{Hu:2003}
Hu, W., Hedman, M., \& Zaldarriaga, M. 2003, \prd, 67, 043004

\bibitem[{Hu \& Okamoto(2002)}]{Hu:2001kj}
Hu, W., \& Okamoto, T. 2002, \apj, 574, 566

\bibitem[{Kesden {et~al.}(2002)Kesden, Cooray, \& Kamionkowski}]{Kesden:2002}
Kesden, M.~H., Cooray, A., \& Kamionkowski, M. 2002, \prl, 89, 011304

\bibitem[{Kesden {et~al.}(2003)Kesden, Cooray, \& Kamionkowski}]{Kesden:2003}
---. 2003, \prd, 67, 123507

\bibitem[{Kirk {et~al.}(2015)Kirk, Omori, {et~al.}}]{Kirk:2015}
Kirk, D., Omori, Y., {et~al.} 2015, arXiv:1512.04535

\bibitem[{Knox \& Song(2002)}]{Knox:2002}
Knox, L., \& Song, Y.-S. 2002, \prl, 89, 011303

\bibitem[{Lee {et~al.}(2015)Lee, Liu, \& Ng}]{Lee:2015}
Lee, S., Liu, G.-C., \& Ng, K.-W. 2015, Phys. Lett. B, 746, 406

\bibitem[{Lewis(2005)}]{Lewis:2005}
Lewis, A. 2005, \prd, 71, 083008

\bibitem[{Lewis \& Challinor(2006)}]{Lewis:2006fu}
Lewis, A., \& Challinor, A. 2006, Phys. Rep., 429, 1

\bibitem[{Lewis {et~al.}(2011)Lewis, Challinor, \& Hanson}]{Lewis:2011}
Lewis, A., Challinor, A., \& Hanson, D. 2011, \jcap, 03, 018

\bibitem[{{LSST Dark Energy Science Collaboration}(2012)}]{LSST}
{LSST Dark Energy Science Collaboration}. 2012, arXiv:1211.0310

\bibitem[{Marian \& Bernstein(2007)}]{Marian:2007}
Marian, L., \& Bernstein, G.~M. 2007, \prd, 76, 123009

\bibitem[{Miller {et~al.}(2008)Miller, Shimon, \& Keating}]{Miller:2008}
Miller, N.~J., Shimon, M., \& Keating, B.~G. 2008, \prd, 79, 063008

\bibitem[{Namikawa {et~al.}(2013)Namikawa, Hanson, \&
  Takahashi}]{Namikawa:2012pe}
Namikawa, T., Hanson, D., \& Takahashi, R. 2013, \mnras, 431, 609

\bibitem[{Namikawa {et~al.}(2010)Namikawa, Saito, \& Taruya}]{Namikawa:2011}
Namikawa, T., Saito, S., \& Taruya, A. 2010, \jcap, 12, 027

\bibitem[{Namikawa \& Takahashi(2013)}]{Namikawa:2013}
Namikawa, T., \& Takahashi, R. 2013, \mnras, 438, 1507

\bibitem[{Namikawa {et~al.}(2016)Namikawa, Yamauchi, Sherwin, \&
  Nagata}]{Namikawa:2016}
Namikawa, T., Yamauchi, D., Sherwin, D., \& Nagata, R. 2016, \prd, 93, 043527

\bibitem[{Namikawa {et~al.}(2012)Namikawa, Yamauchi, \&
  Taruya}]{Namikawa:2011cs}
Namikawa, T., Yamauchi, D., \& Taruya, A. 2012, \jcap, 1201, 007

\bibitem[{Pan \& Knox(2015)}]{Pan:2015}
Pan, Z., \& Knox, L. 2015, \mnras, 454, 3200

\bibitem[{Pearson {et~al.}(2014)Pearson, Sherwin, \& Lewis}]{Pearson:2014}
Pearson, R., Sherwin, B., \& Lewis, A. 2014, \prd, 90, 023539

\bibitem[{{\textit{Planck} Collaboration}(2014{\natexlab{a}})}]{P13:phi}
{\textit{Planck} Collaboration}. 2014{\natexlab{a}}, \aap, 571, A17

\bibitem[{{\textit{Planck} Collaboration}(2014{\natexlab{b}})}]{P13:CIBxphi}
---. 2014{\natexlab{b}}, \aap, A18

\bibitem[{{\textit{Planck} Collaboration}(2014{\natexlab{c}})}]{P13:CIB}
---. 2014{\natexlab{c}}, \aap, A30

\bibitem[{{\textit{Planck} Collaboration}(2015)}]{P15:phi}
---. 2015, arXiv:1502.01591

\bibitem[{Pogosian \& Wyman(2008)}]{Pogosian:2008}
Pogosian, L., \& Wyman, M. 2008, \prd, 77, 083509

\bibitem[{{\textsc{Polarbear}
  Collaboration}(2014{\natexlab{a}})}]{PB14:phixCIB}
{\textsc{Polarbear} Collaboration}. 2014{\natexlab{a}}, \prl, 112, 131302

\bibitem[{{\textsc{Polarbear} Collaboration}(2014{\natexlab{b}})}]{PB14:phi}
---. 2014{\natexlab{b}}, \prl, 113, 021301

\bibitem[{{\textsc{Polarbear} Collaboration}(2015)}]{PB15:rot}
---. 2015, \prd, 92, 123509

\bibitem[{Pospelov {et~al.}(2009)Pospelov, Ritz, \& Skordis}]{Pospelov:2009}
Pospelov, M., Ritz, A., \& Skordis, C. 2009, \prl, 103, 051302

\bibitem[{Pratten \& Lewis(2016)}]{Pratten:2016}
Pratten, G., \& Lewis, A. 2016, arXiv:1605.05662

\bibitem[{Regan {et~al.}(2010)Regan, Shellard, \& Fergusson}]{Regan:2010}
Regan, D., Shellard, E., \& Fergusson, J. 2010, \prd, 82, 023520

\bibitem[{Saga {et~al.}(2015)Saga, Yamauchi, \& Ichiki}]{Saga:2015}
Saga, S., Yamauchi, D., \& Ichiki, K. 2015, \prd, 92, 063533

\bibitem[{Seljak \& Hirata(2004)}]{Seljak:2003pn}
Seljak, U., \& Hirata, C.~M. 2004, \prd, 69, 043005

\bibitem[{Seljak \& Slosar(2006)}]{Seljak:2006}
Seljak, U., \& Slosar, A. 2006, \prd, 74, 063523

\bibitem[{Shaw \& Lewis(2010)}]{Shaw:2010}
Shaw, J.~R., \& Lewis, A. 2010, \prd, 81, 043517

\bibitem[{Sherwin \& Schmittfull(2015)}]{Sherwin:2015}
Sherwin, B.~D., \& Schmittfull, M. 2015, \prd, 92, 043005

\bibitem[{Sigurdson \& Cooray(2005)}]{Sigurdson:2005cp}
Sigurdson, K., \& Cooray, A. 2005, \prl, 95, 211303

\bibitem[{Simard {et~al.}(2015)Simard, Hanson, \& Holder}]{Simard:2015}
Simard, G., Hanson, D., \& Holder, G. 2015, \apj, 807, 166

\bibitem[{Smith {et~al.}(2007)Smith, Zahn, \& Dore}]{Smith07}
Smith, K.~M., Zahn, O., \& Dore, O. 2007, \prd, 76, 043510

\bibitem[{Smith {et~al.}(2012)}]{Smith:2010gu}
Smith, K.~M., {et~al.} 2012, \jcap, 1206, 014

\bibitem[{Story {et~al.}(2015)}]{Story:2014hni}
Story, K.~T., {et~al.} 2015, \apj, 810, 50

\bibitem[{Su {et~al.}(2009)Su, Yadav, \& Zaldarriaga}]{Su:2009}
Su, M., Yadav, A.~P.~S., \& Zaldarriaga, M. 2009, \prd, 79, 123002

\bibitem[{{The Dark Energy Survey Collaboration}(2016)}]{DES}
{The Dark Energy Survey Collaboration}. 2016, \mnras, 1601.00329

\bibitem[{{The DESI Collaboration}(2013)}]{DESI}
{The DESI Collaboration}. 2013, arXiv:1308.0847

\bibitem[{Tolan(2014)}]{Tolan:PhD}
Tolan, J.~E. 2014, PhD thesis

\bibitem[{van Engelen {et~al.}(2012)}]{vanEngelen:2012}
van Engelen, A., {et~al.} 2012, \apj, 756, 142

\bibitem[{van Engelen {et~al.}(2015)}]{vanEngelen:2014zlh}
---. 2015, \apj, 808, 9

\bibitem[{Wu {et~al.}(2014)Wu, Errard, Dvorkin, Kuo, Lee, MacDonald, Slosar, \&
  Zahn}]{Wu:2014}
Wu, W. L.~K., Errard, J., Dvorkin, C., {et~al.} 2014, \apj, 788, 19

\bibitem[{Zaldarriaga \& Seljak(1998)}]{Zaldarriaga:1998ar}
Zaldarriaga, M., \& Seljak, U. 1998, \prd, 58, 023003

\end{thebibliography}

\end{document}